\documentclass{emulateapj}
\usepackage{amsmath}
\usepackage{amssymb}
\usepackage{latexsym}
\usepackage{textcomp}
\usepackage{graphicx}
\usepackage{subfigure}
\usepackage{epsfig}
\usepackage{natbib}

\newcommand{\ha}{\textrm{H}\ensuremath{\alpha}}
\newcommand{\hb}{\textrm{H}\ensuremath{\beta}}

\newcommand{\nii}{[\textrm{N}~\textsc{ii}]}
\newcommand{\oiii}{[\textrm{O}~\textsc{iii}]}
\newcommand{\sii}{[\textrm{S}~\textsc{ii}]}

\newcommand{\oiiilam}{[\textrm{O}~\textsc{iii}]~\ensuremath{\lambda5007}}
\newcommand{\niilam}{[\textrm{N}~\textsc{ii}]~\ensuremath{\lambda6584}} 

\newcommand{\oilam}{[\textrm{O}~\textsc{i}]~\ensuremath{\lambda6300}}
\newcommand{\siilam}{[\textrm{S}~\textsc{ii}]~\ensuremath{\lambda\lambda6717,6731}}

\newcommand{\oiiihb}{[\textrm{O}~\textsc{iii}]/\textrm{H}\ensuremath{\beta}}

\newcommand{\ergs}{\textrm{erg\,s$^{-1}$}}
\newcommand{\Loiii}{$L_{[\textrm{O}~\textsc{iii}]}$}

\newcommand{\Lbol}{$L_{\textrm{bol}}{\textrm{(AGN)}}$}
\newcommand{\mipscol}{$S_{24}/S_{70}$}

\newcommand{\as}{\mbox{\arcsec}}

\slugcomment{Accepted for publication in the ApJ}

\shorttitle{AGN in Star-Forming Galaxies}
\shortauthors{Juneau et al.}

\begin{document}

\title{Widespread and Hidden Active Galactic Nuclei in Star-Forming Galaxies \\ at Redshift $>$ 0.3}

\author{\sc St\'{e}phanie Juneau\altaffilmark{1,2}}
\altaffiltext{1}{CEA-Saclay, DSM/IRFU/SAp, 91191 Gif-sur-Yvette, France; stephanie.juneau@cea.fr}
\altaffiltext{2}{Steward Observatory, University of Arizona, Tucson, AZ 85721, USA}

\author{\sc Mark Dickinson\altaffilmark{3}}
\altaffiltext{3}{National Optical Astronomy Observatory, 950 North Cherry Avenue, Tucson, AZ 85719, USA}

\author{\sc Fr\'ed\'eric Bournaud\altaffilmark{1}}

\author{\sc David M. Alexander\altaffilmark{4}}
\altaffiltext{4}{Department of Physics, Durham University, Durham DH1 3LE, UK}

\author{\sc Emanuele Daddi\altaffilmark{1}}

\author{\sc James R. Mullaney\altaffilmark{4}}

\author{\sc Benjamin Magnelli\altaffilmark{5}}
\altaffiltext{5}{Max-Planck-Institut f\"ur extraterrestrische Physik, Postfach 1312, 85741, Garching bei M\"unchen, Germany}

\author{\sc Jeyhan S. Kartaltepe\altaffilmark{3}}

\author{\sc Ho Seong Hwang\altaffilmark{6}}
\altaffiltext{6}{Harvard-Smithsonian Center for Astrophysics, 60 Garden Street, Cambridge, MA 02138, USA}

\author{\sc S. P. Willner\altaffilmark{6}}

\author{\sc Alison L. Coil\altaffilmark{7}}
\altaffiltext{7}{Department of Physics, Center for Astrophysics and Space Sciences, University of California, 9500 Gilman Dr., La Jolla, San Diego, CA 92093, USA}

\author{\sc David J. Rosario\altaffilmark{8}}
\altaffiltext{8}{Max-Planck-Institut f\"ur extraterrestrische Physik, Giessenbachstrasse, 85748, Garching bei M\"unchen, Germany}

\author{\sc Jonathan R. Trump\altaffilmark{9}}
\altaffiltext{9}{University of California Observatories/Lick Observatory, University of California, Santa Cruz, CA 95064 USA}

\author{\sc Benjamin J. Weiner\altaffilmark{2}}

\author{\sc Christopher N. A. Willmer\altaffilmark{2}}

\author{\sc Michael C. Cooper\altaffilmark{10}}
\altaffiltext{10}{Center for Galaxy Evolution, Department of Physics and Astronomy, University of California, Irvine, 4129 Frederick Reines Hall, Irvine, CA 92697, USA}

\author{\sc David Elbaz\altaffilmark{1}}

\author{\sc S. M. Faber\altaffilmark{9}}

\author{\sc David T. Frayer \altaffilmark{11}}
\altaffiltext{11}{National Radio Astronomy Observatory, PO Box 2, Green Bank, WV 24944, USA}

\author{\sc Dale D. Kocevski\altaffilmark{9,12}}
%%\altaffiltext{12}{Department of Physics and Astronomy, University of Kentucky, Lexington, KY 40506-0055, USA}

\author{\sc Elise S. Laird\altaffilmark{13}}
\altaffiltext{13}{Astrophysics Group, Imperial College London, Blackett Laboratory, Prince Consort Road, London SW7 2AZ, UK}

\author{\sc Jacqueline A. Monkiewicz \altaffilmark{3,14}}
\altaffiltext{14}{School of Earth and Space Exploration, Arizona State University, Tempe, AZ 85287, USA}

\author{\sc Kirpal Nandra\altaffilmark{15}}
\altaffiltext{15}{Max-Planck-Institut f\"ur extraterrestrische Physik, Giessenbachstrasse, 85748 Garching bei M\"unchen, Germany}

\author{\sc Jeffrey Newman\altaffilmark{16}}
\altaffiltext{16}{University of Pittsburgh, Department of Physics and Astronomy, 401-C Allen Hall, 3941 O'Hara Strett, Pittsburgh, PA 15260, USA}

\author{\sc Samir Salim\altaffilmark{17}}
\altaffiltext{17}{Department of Astronomy, Indiana University, Bloomington, IN 47404, USA; salims@indiana.edu}

\author{\sc Myrto Symeonidis\altaffilmark{18}}
\altaffiltext{18}{University College London, Mullard Space Science Laboratory, Surrey RH5 6NT, UK}

\begin{abstract}
We characterize the incidence of active galactic nuclei (AGNs) in $0.3 < z < 1$ star-forming galaxies 
by applying multi-wavelength AGN diagnostics (X-ray, optical, mid-infrared, radio) to a sample of 
galaxies selected at 70-$\mu$m from the Far-Infrared Deep Extragalactic Legacy 
survey (FIDEL).  Given the depth of FIDEL, we detect ``normal'' galaxies on the {\it specific star formation 
rate} (sSFR) {\it sequence} as well as starbursting systems with elevated sSFR.  
We find an overall high occurrence of AGN of 37$\pm$3\%, 
more than twice as high as in previous 
studies of galaxies with comparable infrared luminosities and redshifts but 
in good agreement with the AGN fraction of nearby ($0.05<z<0.1$) galaxies of similar infrared luminosities.  
The more complete census of AGNs comes from using 
the recently developed Mass-Excitation (MEx) diagnostic diagram.  
This optical diagnostic is also sensitive to X-ray weak AGNs and X-ray absorbed AGNs, and reveals 
that absorbed active nuclei reside almost exclusively in infrared-luminous hosts.  
The fraction of galaxies hosting an AGN appears to be independent of sSFR
and remains elevated both on the sSFR sequence and above.  In contrast, the 
fraction of AGNs that are X-ray absorbed increases substantially with increasing sSFR, possibly due to 
an increased gas fraction and/or gas density in the host galaxies.
\end{abstract}


\section{Introduction}

It has become increasingly clear that we need to reconcile the formation of stars in galaxies with 
the growth of the supermassive black holes at their centers for a complete picture of galaxy evolution.  
Theoretically, active galactic nuclei (AGNs) are invoked as a means to quench star formation 
in galaxies thereby explaining the presence of massive red galaxies at recent epoch.  Without 
this so-called AGN feedback, both semi-analytic models and cosmological simulations tend to 
overproduce massive blue galaxies \citep[e.g.,][]{cro06,gab11}.
Together with the tight relation between supermassive black hole (SMBH) mass and bulge mass \citep{mag98,fer00,trema02}, 
this leads to a picture of co-evolution between galaxy stellar content 
and central SMBH \citep[but see][for an alternative interpretation]{jah11}.  
However, observational evidence of a physical connection between star formation and AGN remains 
sparse and mostly indirect, especially during the growth phase (pre-quenching).  

Different triggers of SMBH growth have been proposed: 
major galaxy mergers \citep{san88,dim05,hop06}, large-scale disk instabilities \citep{bou11}, and a secular 
growth where SMBH growth is unrelated to the star formation rate (SFR) of galaxies \citep{sha10,cis11,mul11b}.  Some of these 
studies favor hybrid models where the high-luminosity end (quasar regime, with $L_{AGN}>10^{44}~$erg~s$^{-1}$) 
is dictated by major galaxy mergers while the lower-luminosity AGNs follow a secular evolution 
\citep[e.g.,][]{sha10,lut10,ros12}. 
However, the role of AGNs and their underlying physical connection with host galaxies remain 
uncertain.  %%For instance, 
Most studies of the host galaxy morphologies of high-redshift ($>0.5$) AGNs, 
which have tried to distinguish the relative importance of major merger and secular growth, have relied on X-ray 
selected samples of AGN \citep{gab09,cis11,sch11,koc11}.  The main drawback of this approach as 
a test of the major merger hypothesis is the potential insensitivity to a key phase of AGN growth -- 
when the SMBH is expected to be deeply buried in the gas-rich center of the merging system \citep{san88}.  
During that buried phase, X-ray emission from the AGN may be mostly absorbed by intervening material 
with high column densities and thus may be undetected in even the deepest X-ray surveys currently 
available.  Thus, a real test of this connection and of the AGN content of actively star-forming 
galaxies requires the identification of both absorbed and unabsorbed AGNs.

Nebular emission lines tracing the narrow line regions (NLRs) surrounding AGNs are not subject to the same 
small-scale obscuration as X-rays owing to their much larger physical extent (100's of pc to a few kpc 
scales).  %%Furthermore, 
Indeed, the \oiiilam\ luminosity combined with hard (2$-$10~keV) X-ray luminosity has been used as a 
Compton-thickness parameter in order to infer the X-ray absorption \citep[][hereafter J11; also see Appendix~\ref{app:Xabsor}]{mul94,bas99,hec05,jun11}.  
AGNs identified at visible wavelengths tend to either have a direct view of the nuclear region 
with low-level obscuration by dust in the host galaxy (type 1 AGNs, recognized by broad emission lines)
or a large obscuration of the nuclear region along the line-of-sight combined with a lesser obscuration
of the narrow-line regions (type 2 AGNs, lacking broad emission lines\footnote{Additionally, some 
type 2 (narrow lines only) AGNs may be radiatively inefficient and simply lack broad line regions \citep[e.g.][]{tru11}
but these systems are intrinsically weaker and presumably have less effect on their host galaxies.}).

In extreme cases where both X-ray emission and optical line emission are completely obscured, 
infrared light may %%provide clues
reveal an AGN by showing reprocessed 
thermal emission from AGN-heated dust grains.  Compared to the bulk of stellar light, 
the more energetic radiation field from an AGN heats the dust to higher
temperatures, adding a hot dust component with thermal emission at shorter MIR wavelengths 
\citep{elv94,lac04,ste05,mul11a}.
This extra hot dust alters the mid-IR spectral energy distribution (SED) by 
producing an excess between the usual stellar bump at $\sim1.6\,\mu$m and the stellar-heated 
dust emission at $>10\,\mu$m.  
Broad-band MIR observations such as those from {\it Spitzer}/IRAC can probe that 
feature \citep{lac04,ste05,don07,don12}.  Not all AGNs will exhibit MIR signatures because of the required
geometry and dust content.  Also, intrinsically weaker AGN occurring alongside star 
formation tend not to dominate the IR SED of their host galaxies \citep{bar06}.

Overall, it is clear that no single diagnostic can achieve a complete census of AGNs  
for all galaxies.  As such, combining selections at multiple wavelengths 
may be the key to achieving both a better sampling and a better understanding of AGNs 
and their connection with star formation activity.

In this paper, we characterize the incidence of AGN in star-forming galaxies 
at intermediate redshift ($0.3<z<1$).  
The multi-wavelength dataset and galaxy samples are described 
in Sections~\ref{sec:dataset} and \ref{sec:sample}, respectively. 
The AGN diagnostics are introduced in Section~\ref{sec:diag}.  
Section~\ref{sec:results} contains the results regarding 
the AGN fraction among star-forming FIR-selected galaxies (\S\ref{sec:agnId}); 
the occurrence of active nuclei as a function of infrared luminosity (\S\ref{sec:irAGN}) 
and as a function of specific star formation rate (sSFRs; \S\ref{sec:sSFR}); 
the absorption of AGN in relation to host galaxies (\S\ref{sec:absorb}); 
and the influence of AGN emission on mid-to-far infrared color, probing the dust 
temperature on the warm side of the infrared SED (\S\ref{sec:ir_emline}).  
The caveats are discussed in Section~\ref{sec:caveats}, followed by the possible 
connections between AGN obscuration and host galaxies (Section~\ref{sec:linkHost}), the properties of the 
AGNs selected via different wavelength regimes (Section~\ref{sec:prop}), and possible 
physical interpretations of the triggering mechanisms of AGN in Section~\ref{sec:origin}.
Lastly, we summarize our findings in Section~\ref{summ}. We assume a flat cosmology with 
$\Omega_m = 0.3$, $\Omega_{\Lambda} = 0.7$, and $h = 0.7$ throughout, and a \citet{cha03} 
initial mass function (IMF) when deriving stellar masses and SFRs.

%-----------------------------------------------------------------------------
%\section{Galaxy Samples}\label{sec:sample}

\section{Multiwavelength Dataset}\label{sec:dataset}

The primary intermediate-redshift galaxy sample is based on observations from the 
Great Observatories Origins Deep Survey\footnote{http://www.stsci.edu/science/goods/} 
(GOODS) and the All-wavelength Extended Groth strip International 
Survey\footnote{http://aegis.ucolick.org/} (AEGIS), specifically in the GOODS-North and the Extended Groth Strip (EGS) fields.  
%The datasets used in this work are described below.

\subsection{Infrared Photometry}

Both fields were observed during the Far-Infrared Deep Extragalactic 
Legacy survey \citep[FIDEL;][]{dic07}, yielding sensitive {\it Spitzer} observations 
at 24$\mu$m and 70$\mu$m (respective $3\sigma$ limiting fluxes of $\sim$20$\mu$Jy and 2.5~mJy).  
MIPS photometry was obtained using a guided extraction method, where images at shorter 
wavelength, less subject to confusion, are used to build a set of prior positions 
to fit for the same sources at longer wavelengths.  In this case, IRAC 3.6\,$\mu$m images 
were used to select priors which were then fit simultaneously at 24$\mu$m.  
Next, sources with a 24$\mu$m detection were fit at 70$\mu$m.  The data and 
method are described in more detail by \citet{mag09,mag11}.  Using Monte Carlo simulations, 
these authors found that this method can deblend sources that are at 
least 0.5 $\times$ FWHM apart ($\sim9$\as\ for MIPS 70$\mu$m), and they quantified 
the systematic uncertainties using the difference between the extracted flux 
of simulated sources to the real input flux \citep{mag09}.  
The latest catalogs in GOODS \citep{mag11} include two estimates of the 
uncertainties: the systematic uncertainties quantified from the Monte Carlo simulations, 
as well as local background noise uncertainties (from the residual maps).  
We use the maximum of the two values for each galaxy. 

{\it Spitzer}/IRAC photometry is available in all four channels 
\citep[3.6, 4.5, 5.8 and 8.0$\mu$m; available through the {\it Spitzer} Science Center for GOODS-N and from][for EGS]{bar08}.
In what follows, IRAC photometry is used to calculate rest-frame K-band magnitudes, which 
are in turn used to estimate stellar masses when the latter are not available through SED fitting.  
In addition, IRAC photometry is used for mid-infrared color-color AGN diagnostics 
(\S\ref{sec:irac}).

\subsection{Optical Spectra}\label{sec:spec}

As in the work by \citet[][hereafter J11]{jun11}, the GOODS-N optical spectra are drawn from the Team Keck Redshift 
Survey\footnote{http://tkserver.keck.hawaii.edu/tksurvey/} 
\citep[TKRS][]{wir04}, whereas the EGS spectra come from the DEEP2 Galaxy 
Redshift Survey \citep[hereafter DEEP2;][]{dav03,dav07,new12}.  However, we augment 
the sample with spectra obtained during the DEEP3 campaign 
including galaxies in GOODS-N \citep{coo11} and EGS \citep[][Cooper et al., in prep.]{coo12}.

All sets of observations were obtained with Keck/DEIMOS and 
reduced with the same pipeline\footnote{http://deep.berkeley.edu/spec2d/, developed by the DEEP2 team at the University of California-Berkeley} \citep{coo12,new12}.
However, their spectral resolution and spectral range differ due to the use of 
different gratings (600 line~mm$^{-1}$ for TKRS and DEEP3, and 1200 line~mm$^{-1}$ for DEEP2).
The TKRS and DEEP3 resolution is 4\AA\ FWHM over the wavelength range 5500$-$9800\AA, whereas
DEEP2 spectra have a resolution of 2\AA\ FWHM with a wavelength coverage of 
6500$-$9100\AA.

Emission line fluxes are measured as described in J11, by fitting a Gaussian 
curve to each line individually and integrating the flux over $\pm2.5\sigma$, 
where $\sigma$ is the Gaussian width (=FWHM/2.355).  
Through this paper, we are only concerned with the \hb/\oiiilam\ flux ratio 
and with the \oiiilam\ luminosity.  The \hb\ fluxes are corrected for underlying 
Balmer absorption either by using a population synthesis model fit to the continuum 
(with \citet{bru03} models when the median S/N per pixel is greater than 3), or 
by applying the median equivalent width of $2.8\pm0.9$\AA\ found in J11 to correct 
for Balmer absorption for spectra with low S/N.

\subsection{X-ray Data}

We take advantage of the fact that the Chandra X-ray coverage is very deep: 
2~Ms in GOODS-N \citep{ale03}, 200~ks \citep{lai09,nan05}, and 800~ks in the EGS 
\citep[Laird et al., in prep.; reduced in a similar fashion as][]{lai09}.  
In the EGS, we favor the 800~ks X-ray fluxes when available, and use the wider area 
200~ks observations otherwise. 
The shallower EGS X-ray data (200~ks) are sufficient to detect moderate luminosity X-ray AGNs 
($L_{2-10keV} > 10^{42}~{\rm erg~s^{-1}}$) out to $z \sim 1$.  
Furthermore, we can detect fainter X-ray galaxies (mostly starbursts) in GOODS-N and in 
the 800~ks portion of the EGS.

X-ray observations are primarily used for AGN identification (Section~\ref{sec:Xray}).
Furthermore, they are used to infer X-ray absorption by comparing the hard 
X-ray luminosity ($L_{2-10keV}$, calculated as in J11) to the \oiiilam\ narrow-line 
region luminosity ($L_{\oiii}$), a more isotropic tracer of the intrinsic AGN 
luminosity than X-rays. For that purpose, we derive X-ray upper limits for \oiii-selected 
AGNs.  Upper limits were derived as described by \citet{ale03} in the GOODS-N field, 
and estimated from the sensitivity maps in the EGS as described by J11.  

%\subsection{Radio 1.4~Gz Data}
%
%While the radio method 
%

\subsection{AGN Bolometric Luminosity}\label{sec:Lbol}

Bolometric AGN luminosities are estimated using \Lbol$=10^3\times$\Loiii.  
The bolometric conversion factor to $L_{\rm 2-10~keV}$ has been calculated to range 
from 10$-$70 depending on Eddington ratio and/or X-ray luminosity \citep{vas07,vas09}.
Here we assume \Lbol$=25\times L_{2-10~keV}$ and another factor of 40 to convert from \oiii\ to 
unabsorbed X-ray (2$-$10\,keV) luminosity (the value found by \citet{hec05} for unobscured AGNs).
%In zero cases with an X-ray detection but no values of \Loiii, we instead use $L_{\rm 2-10~keV}$ 
%to derive \Lbol.

The AGN bolometric luminosities can be converted to a BH mass accretion rate with the 
following equation (Equation~1 from \citet{ale12}):
\begin{equation}
\dot{M}_{BH} [\rm{M_{\sun}\,yr^{-1}}] = 0.15  \left(\frac{0.1}{\epsilon}\right) \left(\frac{L_{bol}}{10^{45}\,\rm{erg\,s^{-1}}}\right)
\end{equation}\label{eq:BHAR}
where $\epsilon$ is the efficiency of conversion of mass into energy.  We assume a value 
$\epsilon=0.1$ in our calculations \citep{mar04,mer04}.

\subsection{Infrared Luminosity and Star Formation Rates}\label{sec:sfr}

Total infrared luminosities ($8<\lambda<1000\mu$m) were obtained by fitting the 
70~$\mu$m flux density and redshift with templates from \citet[][hereafter CE01]{cha01}.  
Even when used with only data at much shorter wavelengths (e.g., MIPS~24$\mu$m) 
than the peak of typical infrared SEDs ($\sim$60-100~$\mu$m), these templates 
have been shown to give consistent results with direct integration over the longer wavelength 
{\it Herschel} bands that probe the peak of the FIR SED in galaxies \citep[out to $z\sim1.5$;][]{elb10}.  

Because we are interested in the IR luminosity originating from star formation only, observations at 24\,$\mu$m 
or shorter wavelengths were not included in the fit.  Relative to 70\,$\mu$m, they risk to have 
a larger contamination from AGN-heated dust when galaxy nuclei are active (see Section~\ref{sec:caveats}). 
Using a contaminated 24~$\mu$m measurement would lead to an overestimate of the IR luminosity due to 
stellar heated dust and of the SFR.  %% move some of the CAVEAT discussion here
It is possible that even observations at 70\,$\mu$m have an AGN contribution but the effect will 
be less important than at shorter wavelengths.

Total infrared luminosities were converted to SFRs using the 
relation from \citet{ken98} with a additional conversion factor of 0.66 from a \citet{sal55} 
to a \citet{cha03} IMF.  
% The dependence on the IMF is smaller when 
% using specific star formation rates (sSFR $\equiv$ SFR/$M_{\star}$) than when using stellar 
% masses or SFRs individually. 

%-----------------------------------------------------------------------------
\section{Galaxy Samples}\label{sec:sample}

\subsection{Intermediate-Redshift Samples}\label{sec:selec}

We start from a parent sample of spectroscopically selected galaxies at $z>0.3$ 
with a spectral coverage that includes the expected wavelengths of the \hb\ and 
\oiiilam\ emission lines.  The resulting redshift range is approximately\footnote{In detail, there is 
some variation in the wavelength range of individual spectra 
due to the instrumental set-up (as described in Section~\ref{sec:spec}) and due to the 
exact position of the slits on the mask.} $0.3<z<1$. We will refer to this parent 
sample of 9435 galaxies as the Inter/All sample (where Inter stands for intermediate redshift).  

The Inter/FIR subsample is further defined to have a MIPS 70$\mu$m detection 
with S/N$>3$, yielding a sample of 270 galaxies at $0.3<z<1$ 
(58 in GOODS-N; 212 in EGS). Galaxies in the Inter/FIR sample all 
have a robust 24\,$\mu$m detection (S/N$>8$) given the much greater sensitivity 
of the 24\,$\mu$m observations relative to the 70\,$\mu$m ones.  
The Inter/FIR sample includes 94\% of MIPS 70 galaxies with a spectroscopic 
redshift $0.3<z<0.8$. This percentage decreases to 82\% over the redshift range $0.3<z<1$ 
because the limited spectral range of the DEEP2 setup prevents the coverage of 
the \hb\ and \oiii\ lines beyond $z>0.8$.

Among the Inter/FIR sample, 145 (54\%) galaxies 
have valid emission line measurements with a 3$\sigma$ detection for both the \hb\ 
and \oiiilam\ lines, while 62 (23\%) galaxies have one detection and one upper limit.  
This results in 207 (77\%) galaxies with a constraint on their \oiiilam/\hb\ 
flux ratios.  The remaining galaxies have S/N$<3$ for both emission lines.  

% to have lettered footnotes with deluxetable
\renewcommand{\thefootnote}{\alph{footnote}}

% TABLE
%\begin{deluxetable}{lrcl}
\begin{deluxetable*}{lrcl}
\tabletypesize{\scriptsize}
%\rotate
\tablecolumns{4}
\tablewidth{0pc}
\tablecaption{Definition of Galaxy and AGN Samples\label{tab:samples}}
\tablehead{
   \colhead{Sample}  &  \colhead{Number}  & \colhead{Redshift}  &  \colhead{Criteria}
}
\startdata
\multicolumn{4}{l}{\bf Low-Redshift Comparison Galaxy Sample } \\
Low/FIR       &      4,034   &  $0.05<z<0.1$    &  S/N$>$3 for \hb, \oiiilam, \ha, \niilam\ and \siilam\ emission lines \\
              &              &                  & and detection in IRAS 60~$\mu$m or AKARI 90~$\mu$m  \\
\multicolumn{4}{l}{\bf Intermediate-Redshift Galaxy Samples } \\
Inter/All     &    9,435     &  $0.3<z<1$       &  TKRS and DEEP2/3 spectra that cover the wavelengths of \hb\ and \oiiilam\, \\
              &              &                  &  (from parent samples selected with $R$[AB]$<24.3$ and 24.1, respectively) \\
Inter/FIR     &      270     &  $0.3<z<1$       &  Inter/All galaxies with a valid detection with MIPS 70~$\mu$m (S/N$>$3) \\
\multicolumn{4}{l}{\bf Intermediate-Redshift AGN Selections } \\
X-AGN         &       19     &  $0.3<z<1$       & $L_{2-10\,keV}>10^{42}$\,erg\,s$^{-1}$ and/or HR$>$-0.1 \\
IR-AGN        &       15     &  $0.3<z<1$       & AGN region of IRAC two-color or single-color diagram  \\
MEx-AGN       &       92     &  $0.3<z<1$       & MEx-AGN probabilities   \\
Radio-AGN     &        5     &  $0.3<z<1$       & Radio-excess ($q_{70}<$1.6)   \\
\multicolumn{4}{l}{\bf Intermediate-Redshift AGN Samples } \\
X-Unabsorbed  &       17     &  $0.3<z<1$       & X-ray identified AGNs \tablenotemark{a}  \\
X-Absorbed    &       64     &  $0.3<z<1$       & X-ray unidentified AGNs \tablenotemark{b} 
but intrinsically luminous  \\
              &              &                  & [$\log(L_{\oiii}$[erg\,s$^{-1}$])$>$40.4 or IR-AGN or radio-AGN] \tablenotemark{c}   \\
Weak          &       18     &  $0.3<z<1$       & Fail X-AGN, IR-AGN, radio-AGN selections and with weak \oiiilam\  \\
              &              &                  & [$\log(L_{\oiii}{\rm[erg\,s^{-1}]})<40.4$] \tablenotemark{c}  \\
\enddata
%$^a${Excluding the two most extremely absorbed systems (with $L_{X}/L_{\oiii}<0.3$). } \\
%$^b${Including the two systems with X-ray AGN identification but with $L_{X}/L_{\oiii}<0.3$. }
\footnotetext[1]{Excluding the two most extremely absorbed X-AGNs with log($L_{X}/L_{\oiii}$)$<0.3$. }
\footnotetext[2]{Including the two systems with X-AGN identification but with log($L_{X}/L_{\oiii}$)$<0.3$. }
\footnotetext[3]{Using the same conversion factors as in \S\ref{sec:agnsummary}, $\log(L_{\oiii})=40.4$
  corresponds to $\log(L_{2-10\,keV})=42$ and to $\log(L_{bol})=43.4$ with luminosities in units of erg\,s$^{-1}$.}
%\end{deluxetable}
\end{deluxetable*}

% reset default here
\renewcommand{\thefootnote}{\arabic{footnote}}

\begin{figure}
\epsscale{1.1} \plotone{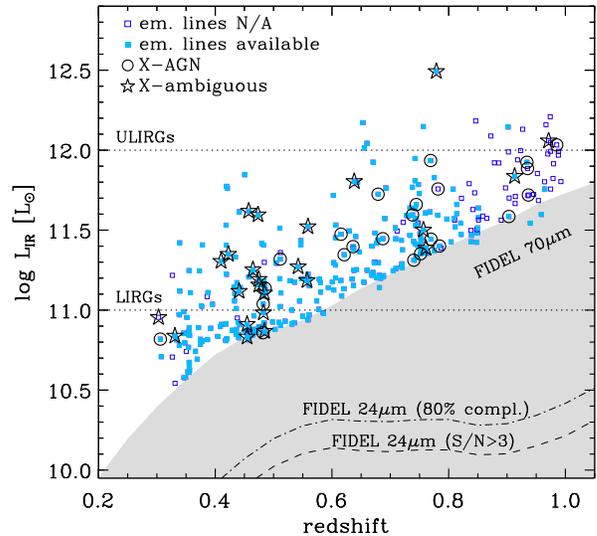}
\caption{Total infrared luminosity as a function of redshift for 70$\mu$m detected galaxies in 
   GOODS-N and the EGS (all symbols). Galaxies with spectral coverage of both the \hb\ and \oiiilam\ 
   emission lines are shown with filled squares (Inter/FIR sample), and the galaxies lacking coverage 
   are shown with open squares.  These features shift beyond the wavelength 
   range of DEEP2 spectra by $z \sim 0.8$, explaining the smaller fraction of galaxies with 
   available emission lines at $0.8<z<1$.  
   The X-ray classification indicates robust X-ray AGNs (open circles) and X-ray ambiguous 
   cases (open star symbols) with faint {\it and} soft X-ray emission like X-ray starbursts.
   We show the $3\sigma$ sensitivity limit of the FIDEL survey at 70$\mu$m with the shaded area
   and the $3\sigma$ limit and 80\% completeness limit of the 24$\mu$m data (dashed and dot-dashed 
   lines, respectively).  All the limits were derived with the CE01 templates.
   }\label{fig:Lir}
\end{figure}

Given the sensitivity of the FIDEL 
survey at 70$\mu$m, the majority (71\%) of the 70$\mu$m-selected galaxies at $0.3<z<1$ 
are luminous IR galaxies (LIRGs; with $L_{IR}(8-1000\mu m) > 10^{11}~L_{\sun}$), 
as shown in Figure~\ref{fig:Lir}.  
These objects are responsible for a large fraction of the global 
SFR density at $z\sim1$ \citep{lef05,mag09} and extend down to the 
typical star-forming galaxies on the {\it main-sequence} of star formation 
\citep{noe07,elb07,elb11}.  Thus, they comprise not only highly starbursting systems, which 
are thought to only contribute $\sim10$\% of the global star formation rate density (SFRD) \citep{rod11,sar12}, but 
also more representative galaxies that contribute to the remaining 90\% of the SFRD.

%---------------------------------------------
% Figure 1 here
%---------------------------------------------
%---------------------------------------------
%---------------------------------------------

\subsection{Low-Redshift Comparison Sample}\label{sec:lozSample}

We build a $z\sim0$ comparison FIR-selected sample from the Sloan Digital Sky Survey (SDSS) DR7 \citep{aba09} 
based on the detection in the IRAS 60$\mu$m or AKARI 90$\mu$m bands as described by \citet{hwa10,hwa10b}. 
Following the procedure in J11, we further select galaxies 
to be at redshift $0.05<z<0.1$ and to have emission line detections with S/N$>3$ 
for all features used in the spectral classification from J11 (\hb, \oiiilam, 
\ha, \niilam, and \siilam) yielding a sample of 4034 galaxies, which we call the  
Low/FIR sample.\footnote{The S/N requirement was relaxed for galaxies 
with SDSS flag $spclass=3$ in order to retrieve broad-line AGNs.  Among 64 
$spclass=3$ objects (1.5\% of Low/FIR sample), 40 have S/N$>3$ in the catalog and the remaining 24 either had a 
lower S/N or were not measured in the MPA/JHU catalog due to the breadth of the lines ($>700~{\rm km~s^{-1}}$).}

The emission line fluxes were obtained from the Value 
Added Catalogs developed by the Max-Planck Institute for Astronomy (Garching) and 
John Hopkins University (MPA/JHU)\footnote{http://www.mpa-garching.mpg.de/SDSS/DR7/}.  
The methodology for measurements of emission line fluxes is described by \citet{tre04}.
We adopt the AGN classification based on the BPT \nii- and \sii-diagrams as 
described by J11, but using the classification scheme from 
\citet{kew06} does not alter our results.

%-----------------------------------------------------------------------------
\section{Multiwavelength AGN Diagnostics}\label{sec:diag}

\subsection{X-ray Diagnostic}\label{sec:Xray}

The X-ray identification of AGNs is based on two criteria: (1) hard X-ray luminosity 
$L_{2-10keV}>10^{42}~{\rm erg~s^{-1}}$, or (2) hardness ratio\footnote{Hardness Ratio 
$\equiv (H-S)/(H+S)$, where $H$ and $S$ are the X-ray count rates in the hard ($2-8$~keV) 
and soft ($0.5-2$~keV) bands, respectively.} $HR>-0.1$, corresponding to an effective photon 
index $\Gamma<1$.  These conditions are similar to the X-ray criteria employed by 
\citet{bau04} based on observed trends that nearby galaxies without AGN do not exceed a 
luminosity of $3\times10^{42}~{\rm erg~s^{-1}}$ in the $0.5-8$~keV band 
\citep[e.g.,][]{fab89}, and that X-ray sources of stellar origin are characterized by 
comparatively soft X-ray emission with photon indices $\Gamma>1$ 
\citep[][also see Figure~2 of \citealt{ale05}]{col04}.  

The 2$-$10\,keV rest-frame luminosities were computed using $k$-correction as described 
by J11 but not corrected for intrinsic absorption within the galaxies.  
Several other studies also adopt such a luminosity threshold when selecting samples of AGNs 
from deep X-ray surveys \citep[e.g.,][]{cis11,sch11,mul11b}.  The addition of a hardness 
criterion in the current study allows us to include weaker and/or more absorbed AGN, 
although only one among 19 X-ray AGN has $L_{2-10keV}<10^{42}~{\rm egs~s^{-1}}$.

X-ray detected sources that do not fulfill the AGN criteria are considered ``X-ray ambiguous'', 
as the X-ray emission could have a stellar origin rather than being due to AGN activity.
However, a faint and soft X-ray emission can also be present in extremely absorbed AGNs if, 
for instance, there is a concurrent X-ray starburst or if the X-ray spectrum is reflection-dominated 
rather than transmission-dominated.  Nearby Compton-thick AGNs are observed to span a broad 
range of spectral indices \citep[see Figure~11 of][]{jun11}.

\subsection{MEx Diagnostic Diagram}\label{sec:mex}

\begin{figure*}
\epsscale{1.15} \plotone{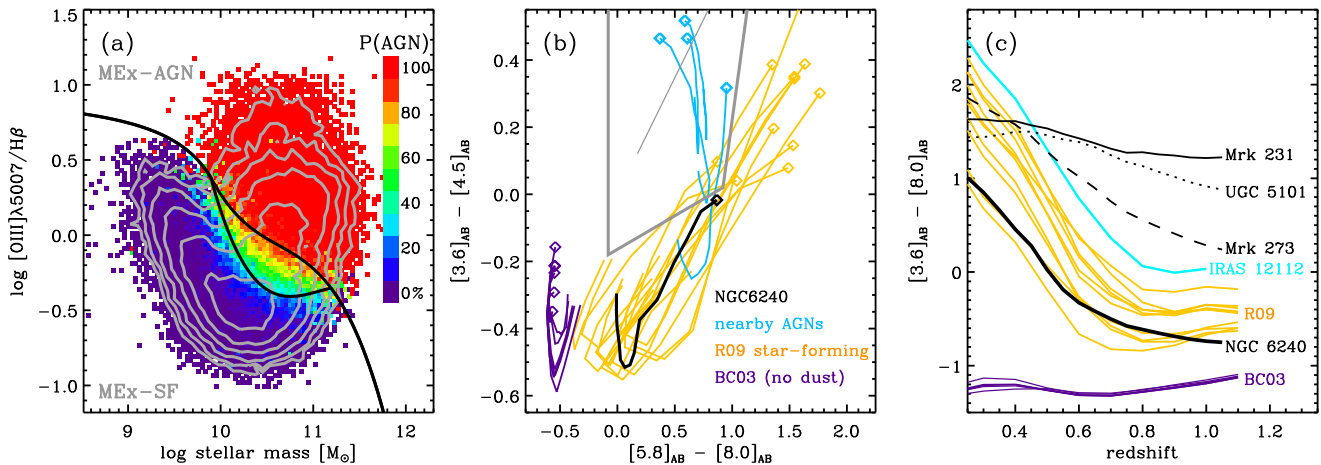}
\caption{
   AGN diagnostic diagrams: 
     (a) The MEx introduced in J11. The empirical 
   curves (solid lines) split the diagrams into galaxy spectral types 
   as labeled, with a MEx-intermediate region between the MEx-AGN and 
   MEx-SF classes.  Logarithmic contours show the number density of the 
   SDSS galaxy sample used to define AGN probabilities (color bar).  
     (b) IRAC color-color diagram from \citet{ste05}.  The AGN selection 
   box (thick solid line) and the power-law locus (thin solid line) 
   are marked in gray. IRAC colors are computed 
   for various templates as a function of redshift from $z=0.3$ 
   (open diamonds) to $z=1$.  Namely, we display the expected colors of 
   nearby star-forming LIRGs and ULIRGs from the templates from  
   \citet[][in yellow]{rie09}, nearby AGNs in light blue, dust-free stellar population 
   models using BC03 templates in purple, and we highlight a 
   nearby Compton-thick double AGN, NGC~6240, with a thick black line. 
     (c) IRAC color between observed channels 1 and 4 as a function of 
   redshift. The cyan curve, which corresponds to the redshift track of IRAS~12112+0305, 
   is the star-forming galaxy template showing the most extreme IRAC colors.
   The remainder of the star-forming galaxy templates from \citet{rie09} 
   (yellow tracks) span a fairly restricted range in color.  Galaxies above 
   that range are suspected to host AGN activity.
   This picture is supported by the projected colors of a 
   few nearby AGNs [black lines: Mrk~231 (solid); UGC~5101 
   (dotted line); and Mrk~273 (dashed line); but 
   also see Compton-thick AGN NGC~6240 (thick solid)].
   Dust-free stellar population models from BC03 define the lower 
   envelope (purple tracks).
}\label{fig:intromex}
\end{figure*}

The second AGN identification method is the MEx diagnostic diagram introduced by J11.
The MEx diagram plots stellar mass versus the \oiiilam/\hb\ flux ratio (Figure~\ref{fig:intromex}a).
Empirical dividing lines indicate spectrally distinct regions of the MEx diagram: optically-identified 
AGN tend to lie above the lines, purely star-forming galaxies below the lines and the 
MEx-intermediate region (between the lines) contains a sharp transition from star-forming to AGN galaxies 
as indicated by the strong gradient from P(AGN)=0\% to P(AGN)=100\%. 
While we use the dividing lines for discussion and visualization, the analysis relies more 
importantly on the AGN probability values\footnote{See https://sites.google.com/site/agndiagnostics/ 
for software to  compute the MEx AGN classification.}.  Those probabilities are determined by using a set of 
priors from the SDSS as described by J11 and correspond, for a given observed galaxy, to the fraction of AGNs 
of any type (Seyfert 2, composite, LINER) relative to the total number of galaxies located within 
the measured uncertainties on the \oiiilam/\hb$-$M$_{*}$ plane.  

The main strengths of the MEx diagram are its applicability to higher redshift relative to 
traditional BPT diagrams\footnote{BPT diagrams require \ha\ and other fairly red emission lines 
(\niilam, \siilam, or \oilam), which redshift out of the visible window at $z>0.4$.} and the use 
of a probabilistic approach to assess the reliability of a given spectral class.  
In the remainder of this work, we use the AGN probability $P(AGN)$ 
to count the number of AGN (e.g., a galaxy with $P(AGN)=30$\% is counted as 0.3 AGN instead 
of 1 AGN) in cases where an AGN is identified only with the MEx method.
The AGN fraction is defined as follows:
\begin{equation}
   {\rm AGN~fraction} = \frac{1}{N} \displaystyle\sum\limits_{i=1}^N P(AGN)_i
\end{equation}\label{eq:frac}
where the AGN probability $P(AGN)$ varies from 0 to 1 and is summed over the number of galaxies $N$ that 
belong to the subsample of interest.  
This approach reduces the risk of contamination by star-forming galaxies into the AGN samples but 
may lead to undestimating the total AGN fraction as composite galaxies, which mostly harbor AGNs \citep[e.g.][]{tro11}, 
tend to have P(AGN)$<$100\%.  
The initial values of $P(AGN)$ are obtained from the 
MEx diagnostic ($P_{MEx}$(AGN)).  However, AGNs identified from an alternative method (X-ray or IR colors) 
are set to have $P(AGN)=1$.  The 68.3\% confidence intervals on fractions are determined with Bayesian binomial 
statistics following the formalism and IDL\footnote{Interactive Data Language} implementation of \citet{cam11}.  

The MEx diagnostic diagram successfully detects AGNs that are missed in even the deepest X-ray surveys.  
J11 showed that at least some of the X-ray undetected galaxies with a probability $>30$\% of hosting 
an AGN contain X-ray absorbed AGN activity.  Indeed, X-ray stacks revealed 
detections in both the soft and hard bands of the {\it Chandra} X-ray observations with a 
flat spectral slope indicating X-ray absorption. 
Although the MEx AGN classification may not always hold on an individual galaxy basis, it was 
validated overall to $z=1$ for average galaxy populations \citep{jun11}, and preliminary studies 
at slightly higher redshifts suggest that the MEx method remains valid on average 
($z \sim 1.5$; Trump et al., submitted).  Furthermore, \citet{mul11b} found that the 
mean stellar mass of X-ray selected AGN hosts did not evolve between $0.5<z<3$.  
This trend suggests that the diagnostic power of the MEx diagram may hold at those higher redshifts 
for the moderate to luminous AGNs ($L_{bol}>10^{43.4}$\,erg\,s$^{-1}$).  
Nevertheless, we limit the analysis to $z<1$ in this work, where the MEx diagnostic can be 
applied with a high level of confidence.

%---------------------------------------------
% Figure 2 here
%---------------------------------------------
%---------------------------------------------

\subsection{IRAC Single-Color and Two-Color Diagnostics}\label{sec:irac}

Infrared AGN identification methods rely on probing AGN-heated dust. 
\citet{lac04} and \citet{ste05} reported that combinations of IRAC colors select 
both obscured and unobscured luminous AGN.  These studies were based on fairly 
shallow IRAC data. \citet{bar06} found that the Lacy and Stern MIR diagnostics, 
when applied to more sensitive IRAC data, 
miss a significant fraction of X-ray selected AGNs and suffer from more contamination from 
non-AGN galaxies.  The latter is caused by the deeper observations probing to higher redshift ($z>1$) 
and making the IRAC colors of starburst and normal star-forming galaxies appear redder and 
moving into the AGN selection box \citep{don07}.

Here, we adopt the IRAC color-color diagram developed by \citet{ste05} because it was shown to 
suffer from less star-forming galaxy contamination than the Lacy diagram when applied to higher-redshift 
and to more sensitive IRAC observations \citep{don07}.  The diagram is shown in Figure~\ref{fig:intromex}(b), 
where AGNs nominally lie in the upper boxed region. We show some tracks of expected mid-infrared colors for various 
galaxy templates.  We do not expect strong contamination by 
star-forming galaxies at $z<1$, but some well-known AGNs are located 
outside of the AGN region when their colors are redshifted.  

Because IRAC color-color diagrams do not include information about the redshift of the galaxies,
there are degeneracies in observed colors from overlapping redshift tracks of different galaxy populations 
\citep[see, e.g.,][]{don07}.  To remove such degeneracies, we use the IRAC [3.6]-[8.0] color as a function 
of redshift.  As shown in Figure~\ref{fig:intromex}(c), infrared-selected star-forming galaxies 
occupy a distinct region from nearby AGN\footnote{The AGN templates were 
constructed by gathering photometry from the NASA Extragalactic Database and interpolating with 
templates (R. Chary, private communication, 2008).},
especially at $z > 0.5$.  
The outermost star-forming galaxy template is for IRAS~12112+0305, which we use to 
separate the IRAC star-forming galaxies (below the cyan curve) from the IRAC AGN candidates (above the cyan curve). 
The change in IRAC [3.6]-[8.0] color with redshift of the infrared selected star-forming galaxies  
is due in part to the aromatic features at rest-frame wavelengths $6.2-8.6\,\mu$m gradually shifting outside of 
the [8.0] band (between $0<z<0.4$) and the stellar bump at rest-frame $\sim1.6\,\mu$m entering the [3.6] band at $0.7<z<1.2$.
When AGN-heated dust dominates the mid-IR emission, the near power-law shape of the continuum causes 
a flatter [3.6]-[8.0] change of color with redshift, as exemplified by Mrk~231 and UGC~5101.  Lastly, 
dust-free stellar populations templates \citep{bru03} set a lower envelope to the observations.  
When galaxies are nearly or completely dust-free,  
the IRAC color corresponds to that of stellar photospheres.

\subsection{Radio AGN Diagnostic}

Radio-excess AGNs can be identified by an excess of radio emission relative to the FIR-radio correlation  \citep[e.g.][]{roy97,don05,del12}, 
which otherwise is very tight for star-forming galaxies and many radio-quiet AGNs \citep[see review by][]{con92}. 
Radio AGN identification was found to reveal a different AGN population with limited overlap with the 
X-ray and IRAC color AGN diagnostics \citep{hic09}.  However, radio AGNs were also found to reside in different host galaxies, 
characterized with, on average, higher stellar masses, stronger clustering, and passive stellar activity compared 
to the hosts of X-ray or IR AGNs \citep{hic09}.  The typically more passive galaxies expected to host radio AGNs 
are unlikely to be detected in the MIPS 70\,$\mu$m observations used to draw our sample.  Therefore, we expect that this category 
of AGNs will not be important for the sample selected here.  % and we will not include this selection in the analysis.

%% Nevertheless, we checked the number of AGNs that could be missed by neglecting the radio AGN population to confirm that their 
%% exclusion does not significantly affect any of the results presented in this Paper.  
In the GOODS-N field, there are deep 
Very Large Array (VLA) 1.4~GHz observations reaching 3.9~$\mu$Jy beam$^{-1}$ rms with a beam size of $\sim$1.\arcsec7 \citep{mor10}.  
These data were used to identify radio-excess AGNs by \citet{del12} for counterparts of 24\,$\mu$m sources which have S/N$>$3 
in the 1.4~GHz observation ($S$(1.4\,GHz) $>$ 12\,$\mu$Jy). 
Infrared SED fitting was used to compute FIR emission integrated over rest-frame $\sim 40-120\,\mu$m in order to 
apply a standard FIR-radio correlation\footnote{This was shown by the authors to be more sensitive than using 
MIPS 24\,$\mu$m as a monochromatic proxy of the FIR with the selection criterion of \citet{don05}: 
$q_{24}\equiv \log (S_{24~\mu m}/S_{1.4~GHz}) < 0$.}. 
Only one galaxy in our GOODS-N subsample is selected as a radio-excess source by \citet{del12}.  
This galaxy is also identified as AGN 
from all the other methods (X-ray AGN, IR AGN, and MEx-AGN) and therefore does not add a new AGN population.  

In the EGS, VLA observations were presented by \citep{ivi07} along with a catalog of sources with $S$(1.4\,GHz) $>$ 50 $\mu$Jy 
($\sim$5$\sigma$) and a $\sim$3.$\arcsec$8 FWHM synthesized beam.  For the 41 radio sources with a match in the Inter/FIR 
sample, we computed $q_{70}\equiv \log (S_{70~\mu m}/S_{1.4~GHz})$ as a proxy for the FIR-radio correlation.  
The median of $q_{70}$=1.9 is consistent to values previously reported in the same redshift range 
\citep[1.9$\pm$0.2 and 1.9$\pm$0.1 from][respectively]{fra06,bour11} and, as expected, 
slightly lower than k-corrected $q_{70}$ values \citep[see][and the compilation in their Table~5]{sar10}.  
Adopting $q_{70}<1.6$ to define radio-excess AGNs yields four AGN candidates\footnote{This cut corresponds 
to a $>$3\,$\sigma$ excess according to the work of \citet{bour11}, and to a 1.5\,$\sigma$ excess according 
to the work of \citet{fra06}.  The significance of the excess may thus depend on the number of galaxies and 
the sample selection.}.  
Two of them are identified as MEx-AGNs (but undetected in X-rays and classified as star-forming on the 
IR diagnostics), while the other two are not identified as AGN by any method.   
One of the unidentified radio AGNs has optical signatures according to \citet{yan11}, who assumed 
a more relaxed contraint on emission line detection (S/N$>$2 instead of S/N$>$3 as in the current work).  

Combining both GOODS-N and EGS fields gives a total of five radio-excess AGN candidates, three of which are 
also identified with alternative methods.  The remaining two radio AGNs make only 0.7\% of the Inter/FIR sample, and 
1.9\% of the AGN subsample.  Therefore, their inclusion or exclusion does not alter the results presented in this Paper, 
but we include them in the analysis whenever the plotted quantities are available.  Namely, the two radio-AGNs 
that are otherwise unidentified will be put in the X-ray absorbed category (Section~\ref{sec:agnsummary}).

%-----------------------------------------------------------------------------
%%%%
%%%% Section 5: RESULTS
%%%%

\section{Results}\label{sec:results}

\subsection{AGN Identification}\label{sec:agnId}

\begin{figure*}
\epsscale{1.15} \plotone{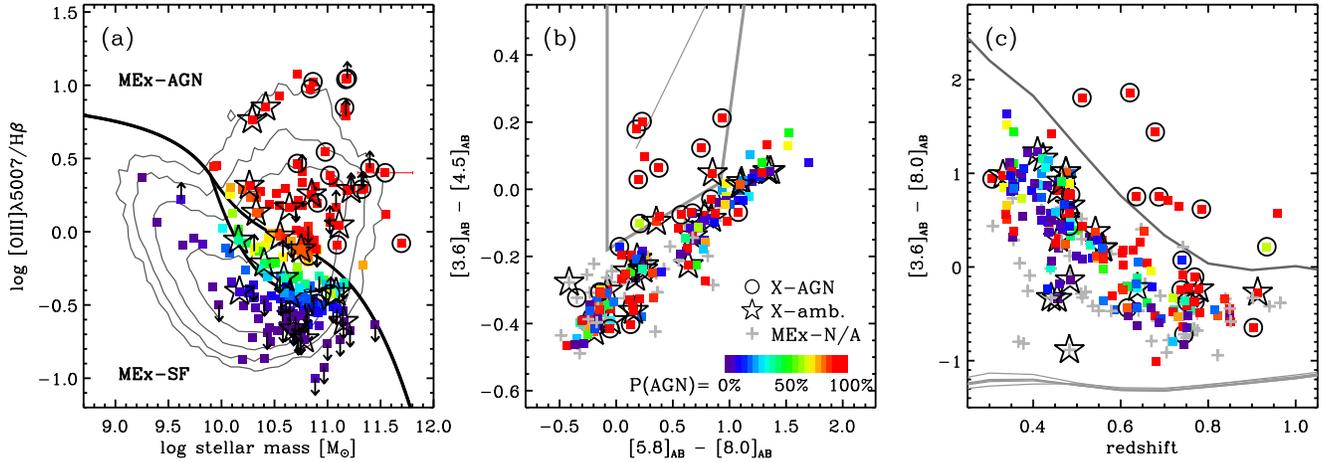}
\caption{
   AGN diagnostic diagrams introduced in Figure~\ref{fig:intromex}: 
   (a) The MEx AGN diagnostic diagram.  The empirical 
   curves (solid lines) split the diagrams into galaxy spectral types 
   as labeled, with a MEx-intermediate region between the MEx-AGN and 
   MEx-SF classes. Contours show the 
   SDSS low-$z$ sample (evenly spaced logarithmically), and the 
   70$\mu$m sample is superimposed with larger symbols keyed to the P(AGN) (color bar) 
   and to the presence of X-ray AGN (circles) or ambiguous X-ray source (possibly X-ray starbursts, 
   star symbols). 
   (b) IRAC color-color diagram from \citet{ste05}.  
   The colored symbols are keyed to the MEx-AGN probabilities (color bar) whereas the gray plus symbols  
   mark galaxies lacking a valid MEx classification due to low S/N \oiiilam\ and \hb\ lines. 
   X-ray identified AGNs (X-ray ambiguous sources) are further marked with a circle (star symbol).  
   Most IRAC AGN candidates in our sample are also recognized as such from the X-ray and/or MEx selection methods.
   (c) IRAC color as a function of redshift. Same plotting symbols as the previous panel.  
   The lower lines are the BC03 tracks for dust-free stellar populations whereas the top gray 
   lines is the template of IRAS~12112, used as an AGN-star formation dividing line on this diagram.  
}\label{fig:agn}
\end{figure*}

The Inter/FIR sample is shown on the MEx diagram in Figure~\ref{fig:agn}(a).  
Most (16/19$=$84\%) of the X-ray AGNs lie in the MEx-AGN and MEx-intermediate regions, as expected.   
In addition, there are 78 AGNs identified from the MEx diagnostic that were missed in 
the X-ray classification, including 14 X-ray ambiguous sources.

%---------------------------------------------
% Figure 3 here
%---------------------------------------------
%---------------------------------------------

In Figure~\ref{fig:agn} (b), we compare IRAC color-color identified AGN with the X-ray 
and optical MEx classifications. 
The X-ray AGNs display a similar distribution as the optically-selected AGN 
on the IRAC color-color plane.  
A few of the IRAC-AGN candidates which are undetected in X-rays are classified as AGNs 
with the optical diagnostic.  Combining both the X-ray and MEx classifications selects
90\% (9/10) 
%%86\% (12/14) 
of the galaxies within the IRAC selection box.  %% The remaining two IRAC AGN candidates 
%% are either genuine (but obscured) AGN or star-forming contaminants.  

The X-ray, MEx and IRAC color-color AGN selection methods are next shown on the IRAC 
single-color diagnostic described in Section~\ref{sec:irac} and displayed in Figure~\ref{fig:agn}(c).
Unsurprisingly, the FIR-selected galaxies mostly follow the tracks of the nearby infrared galaxies from Figure~\ref{fig:intromex}.  
Galaxies hosting X-ray AGNs (black circles) span the whole range of 
IRAC colors covered by the Inter/FIR sample (from non-AGN and up to the most extreme AGN-like color).  The 
X-ray AGNs that are indistinguishable from star-forming galaxies are likely systems where star formation 
dominates at infrared wavelengths even in the presence of nuclear activity \citep[e.g.,][]{ruj11}.  
We note additional AGN candidates that were not selected by the X-ray method (open black 
circles) nor by the IRAC color-color diagram (Panel b).  
In the remainder of this paper, IRAC-selected AGN (or simply IR-AGN) refer to AGN identified from 
either the color-color or the single-color versus redshift method.

\subsection{Summary of AGN Selection and AGN Properties}\label{sec:agnsummary}

\begin{figure}
\epsscale{.9} \plotone{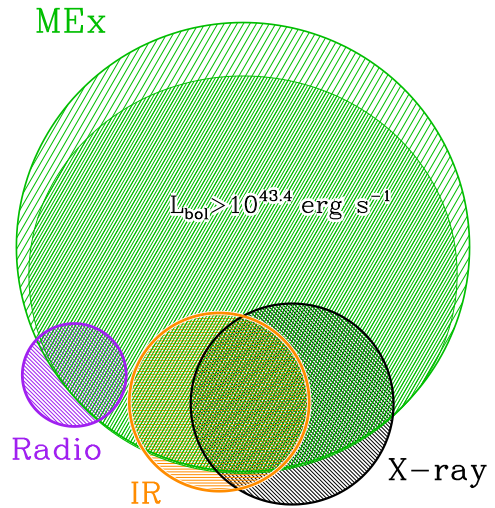}
\caption{
   Venn diagram of the AGN subsamples: 
   MEx-AGN (in green), radio-excess AGN (in purple), IR-AGN (in orange) and X-AGN (in black).  The shaded areas 
   are proportional to the number of galaxies in each category and overlap region.  
   The MEx-AGN sample is further split into bins of AGN bolometric luminosity ($L_{bol}$) as labeled. 
   The luminosity cut at $L_{bol}>10^{43.4}$\,\ergs would only remove one AGN from the IRAC AGN 
   subsample, corresponding to 7\% (1/15) of the area shown in the figure.
      }\label{fig:venn}
\end{figure}

We illustrate the overlap and complementarity between the AGN classification methods 
in Figure~\ref{fig:venn}.  
The overlap between the MEx method and the IR and X-ray methods is greater for the more luminous AGNs, 
as expected because they are easier to detect with any given tracer explored here.  
X-ray selection suffers from ambiguity between faint AGNs and starburst galaxies at low X-ray 
luminosities ($L_{2-10~keV}<10^{42}$\,erg\,s$^{-1}$, or \Lbol$<10^{43.4}$\,erg\,s$^{-1}$) and the 
IRAC selection is mostly sensitive to intrinsically luminous and dusty AGNs 
\citep[$L_{bol}>10^{43}$~erg~s$^{-1}$;][]{hic09,don12}.  
Although optical emission lines can be subject to attenuation by dust in the host galaxy, they remain 
the most sensitive AGN tracer and probe to much lower accretion rates due 
to the detection of relatively faint emission lines in optical spectra (in this work, down to fluxes of 
$\sim 7 \times 10^{-18}$\,erg\,s$^{-1}$\,cm$^{-2}$, corresponding to \Lbol$\sim 10^{42.5}$\,erg\,s$^{-1}$ 
and accretion rates of $5\times10^{-4}$\,M$_{\sun}$\,yr$^{-1}$).  The greater sensitivity 
likely explains the additional AGN populations that are not detected with alternative methods.  
However, there may be cases where the narrow line region emission lines are extinguished by surrounding dust.  
It is important to combine different AGN methods to take advantage of their 
complementarity and achieve a more complete census.  Furthermore, the combination of multiwavelength information 
can distinguish between different AGN sub-populations (or AGN accretion phases).

%---------------------------------------------
% Figure 4 here
%---------------------------------------------
%---------------------------------------------

The different AGN tracers are combined in order to divide the AGN population into three categories 
that are physically meaningful:
\begin{enumerate}
\item{{\bf X-unabsorbed AGN:} X-ray identified AGNs except for two 
systems that have the highest absorption ($N_H>10^{24}$\,cm$^{-2}$, inferred from 
$\log(L_{X}$/\Loiii$)<0.3$; Appendix~\ref{app:Xabsor})\footnote{The remainder may contain moderately absorbed AGN 
as we do not measure the absorbing column density from X-rays.  The latter would be highly uncertain 
due to low number of counts for most objects.}.}
\item{{\bf X-absorbed AGN:} AGNs that are unidentified in X-rays -- plus the two most absorbed systems 
excluded from the X-unabsorbed category -- but which are inferred to be intrinsically luminous due 
to their identification with an IR or radio method or due to having $L_{\oiiilam}>10^{40.4}$\,erg\,s$^{-1}$, 
corresponding to $L_X>10^{42}$\,erg\,s$^{-1}$. These AGNs should thus be detectable with the 
X-ray method in absence of absorption.}  
\item{{\bf Weak AGN:} AGNs which are not identified with X-rays nor IRAC colors and 
which have a faint \oiiilam\ luminosity (\Loiii$<10^{40.4}$\,erg\,s$^{-1}$), implying that 
they would not be recognized as AGNs with the X-ray definition even if they were X-ray 
detected.}
\end{enumerate}
These AGN subsamples are mutually exclusive, and the subsample sizes and definitions 
are summarized in Table~\ref{tab:samples}.  They share interesting similarities and 
differences in their host galaxy and accretion rate properties, and may correspond to 
different phases in a typical SMBH growth cycle (Section~\ref{sec:prop}).

In what follows, we compute the distributions of AGN and star-forming (SF) galaxy subsamples 
by adding the probabilities of each galaxy falling in the category of interest.  The number 
of galaxies per bin is the sum of (1-P(AGN)) for the SF subsample and the sum of P(AGN) for 
the AGN subsamples.  When calculating numbers, P(AGN) is converted as a number from 0 to 1 rather 
than a percentage. Similarly, the probability-weighted means are obtained as follows:
\begin{equation}
   \langle X \rangle = \frac{\displaystyle\sum\limits_{i=1}^N X_i \times P(AGN)_i}{\displaystyle\sum\limits_{i=1}^N P(AGN)_i}
\end{equation}\label{eq:probMean}
where $X$ is the quantity of interest ($M_{\star}$, \Loiii, or \Loiii/$L_{IR}$), and the 
sums are over all the $N$ galaxies in a given subsample.  P(AGN) is replaced with 1-P(AGN) for the 
SF sample.  This methodology takes full advantage of the probabilistic approach of the MEx diagram.

\begin{figure*}
\epsscale{1.15} \plotone{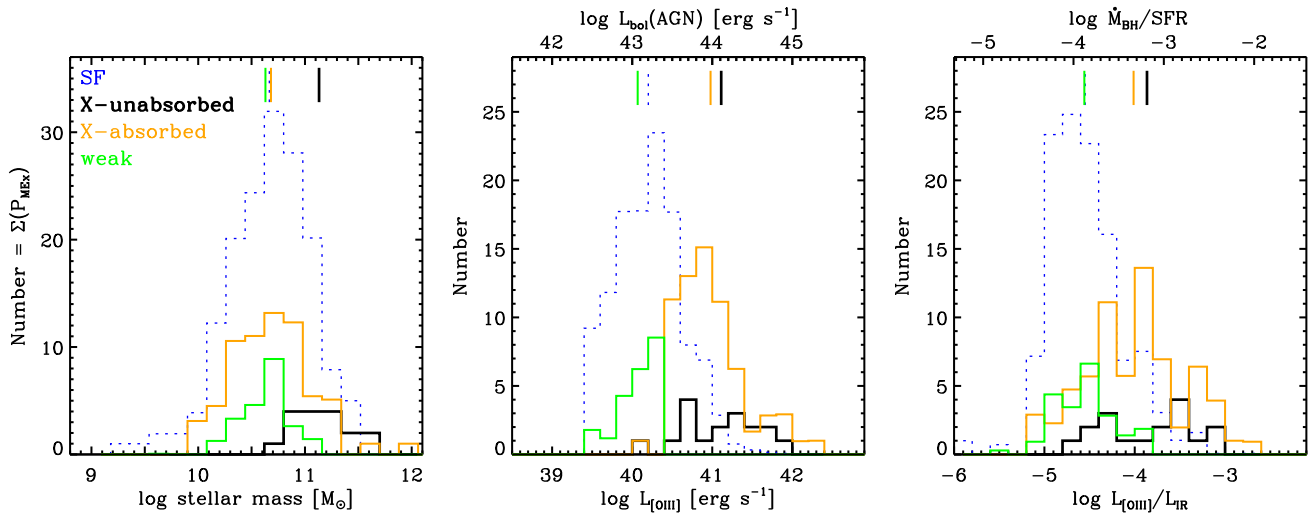}
\caption{
   Distribution of host galaxy stellar masses (left), \oiii\ luminosities (center; corresponding AGN bolometric 
   luminosities on top axis), and \oiii-to-IR luminosity ratios (right; corresponding BHAR 
   to SFR ratios on top axis).
   Those properties are shown for star-forming galaxies (blue dotted line) and 
   the three AGN subsamples : X-unaborbed (black), X-absorbed (orange) and intrinsically weak (green). 
   The physical meaning of $L_{bol}(AGN)$ and BHAR does not apply to the SF subsample (unless some of those 
   galaxies contain unidentified AGNs).  
   The {\it number} of galaxies is computed with $\sum{\rm (1-P(AGN))}$ for the SF subsample and with
   $\sum {\rm P(AGN)}$ for the AGN subsamples 
   (as a reminder, P(AGN)$={\rm P_{MEx}(AGN)}$ for MEx-only AGNs and P(AGN)$=1$ for X-AGNs and IR-AGNs).
   The probability weighted means (Eq.~\ref{eq:probMean}) are indicated with vertical lines.
}\label{fig:distr}
\end{figure*}

As can be seen in Figure~\ref{fig:distr}, X-unabsorbed AGNs are only 
found in the most massive hosts ($\sim10^{10.6-12}$\,M$_{\sun}$) whereas X-absorbed and weak 
AGNs span a broader range of stellar masses ($\sim10^{10-12}$\,M$_{\sun}$) 
and have a similar average stellar mass as the non-AGN galaxies in the Inter/FIR sample 
($\sim10^{10.6}$\,M$_{\sun}$), which is $\sim0.5$~dex lower than the average stellar mass of 
the X-unabsorbed sample.  The apparent lack of X-unabsorbed AGNs in lower mass hosts  
may be due to a combination of AGN evolutionary phase and selection bias (discussed in 
Section~\ref{sec:prop}).

Absorbed and X-unabsorbed AGNs have similar AGN luminosities on average (\Lbol$\sim10^{44}$\,erg\,s$^{-1}$; 
inferred from \oiii; Figure~\ref{fig:distr}) 
and are systematically higher (by 1~dex on average) than the weak AGNs and the SF galaxies.  
The physical meaning of $L_{bol}(AGN)$ does not apply to the SF subsample (unless the galaxies 
contain unidentified AGNs).  
However, the similarity of the \oiii\ luminosities between weak AGNs and SF galaxies (similar means, 
and relatively similar distributions) implies that we are able to distinguish between these two 
populations based on an additional factor and that the \oiii\ luminosity alone is not sufficient 
to discriminate between AGNs and star-forming galaxies.  This may be because \oiii\ may contain 
a contribution from low-metallicity gas ionized by star formation.  This contribution is expected 
to be negligible in metal-rich galaxies \citep{kau03c}, but we test this possible contamination to 
\oiii\ by examining the \oiii-to-IR luminosity ratios.  If the \oiii\ emission traced the SFR instead of the 
AGN luminosity, and assuming that $L_{IR}$ traces the SFR, we would expect similar ratios 
between star forming galaxies and the false AGN candidates.  In the case of genuine BH 
growth with small or no stellar contribution to \oiii, the \oiii-to-IR ratios can be converted as black hole 
accretion rate (BHAR)-to-SFR ratios.  
Absorbed and X-unabsorbed AGNs have similar BHAR-to-SFR ratios, and are systematically higher 
(by ~0.5-0.6 dex) than the weak and SF subsamples indicating an \oiii\ excess emission 
(Figure~\ref{fig:distr}). 
The later two subsamples exhibit similar \oiii-to-IR 
distributions to one another, suggesting that the weak AGNs are not only weak relative to 
the absorbed and X-unabsorbed subsamples, but also relative to the SFR of their hosts.
Some of the weak AGN may represent a distinct AGN regime with possibly radiatively 
inefficient accretion \citep{pta98,ho08,tru11} rather than a simple extension of the same accretion 
process down to lower luminosity \citep[but also see][]{mao07}. 

%---------------------------------------------
% Figure 5 here
%---------------------------------------------
%---------------------------------------------

\subsection{AGN Fraction in Intermediate-Redshift Infrared Galaxies}\label{sec:irAGN}

In order to obtain the global AGN fraction, we combine all of the AGN identification 
methods, namely: the MEx diagnostic diagram, the X-ray classification, and the IRAC single-color and 
two-color diagrams.  
Adopting the procedure from Section~\ref{sec:mex} and Eq.~\ref{eq:frac}, the global AGN fraction 
in $0.3<z<1$ FIR-selected galaxies is 37\%$\pm$3\% (99/270).

\begin{figure*}
\epsscale{1.05} \plotone{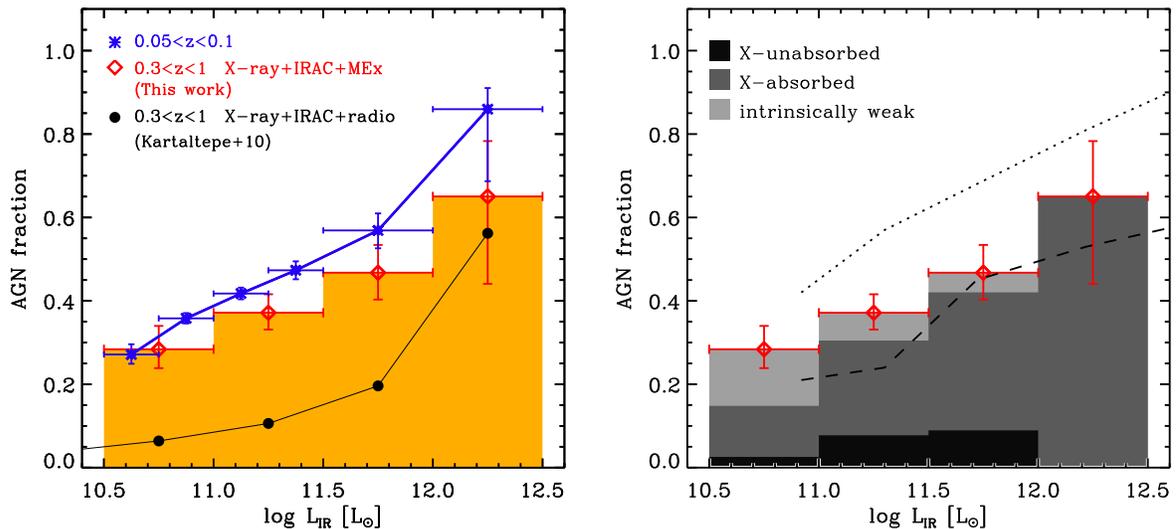}
\caption{ {\bf Left.}
   AGN fraction as a function of $L_{IR}$ in intermediate-redshift FIR-selected 
   galaxies (filled histogram).  
   The AGN fraction rises with increasing $L_{IR}$ in agreement with the trend 
   observed for low-redshift ($0.05<z<0.1$) FIR-selected SDSS galaxies from 
   \citet[shown in blue]{hwa10}.    
   For comparison, we also show values obtained for the 70$\mu$m selected 
   sample from \citet{kar10a}, 
   spanning similar luminosity and redshift ranges (filled black symbols).
   {\bf Right.} 
   Similar to the left-hand panel except that the AGN content is broken down into 
   X-ray unabsorbed, X-ray absorbed, and intrinsically weak systems, as labeled. 
   Interestingly, the overall trend of the AGN fraction 
   against $L_{IR}$ resembles that of the galaxy merger fraction (dashed line) and 
   the combined irregular and merger morphology fraction (dotted line) observed by 
   \citet{kar12} at $z\sim1$ (discussed in Section~\ref{sec:origin}). 
   Horizontal error bars show the bin width and vertical error bars show the 68.3\% 
   confidence intervals for fractions, determined with Bayesian binomial statistics 
   following the formalism of \citet{cam11}.
   }\label{fig:AGNfrac_IR}
\end{figure*}

The AGN fraction increases steadily with $L_{IR}$ and both the increase and the magnitude of the 
AGN fraction are consistent with the values obtained for the low/FIR comparison 
sample (Figure~\ref{fig:AGNfrac_IR}).  The similarity 
between the AGN fractions of the Low/FIR and Inter/FIR samples is consistent with no or very mild evolution in 
the triggering of AGN as a function of IR luminosity between $z\sim1$ and $z\sim0$.

In contrast, the AGN fraction found for the Inter/FIR sample is significantly higher than that of 
similarly selected galaxies in the COSMOS field \citep{kar10a}.  Those authors 
combined X-ray, IRAC power-law, and radio AGN identification methods \citep[also see][]{sym10} 
but lacked a robust optical emission line diagnostic\footnote{\citet{sym10} used a simple 
\oiiilam/\hb\ $>3$ threshold, which only selects AGNs with the most extreme line ratios.  This selection criterion 
would yield only 10 out of 92 (11\%) MEx-AGNs.}.  
Our higher AGN fractions are mostly a result of a more complete AGN selection due to the inclusion of 
X-ray faint and absorbed AGNs from the MEx diagram.  The latter account for 
most of the difference between the samples (right panel of Figure~\ref{fig:AGNfrac_IR}).

%---------------------------------------------
% Figure 6 here
%---------------------------------------------
%---------------------------------------------

There are more X-ray absorbed AGN than X-ray unabsorbed AGN at all IR luminosities but the distributions 
of the infrared luminosities of these two AGN categories are similar. The intrinsically weaker 
AGN tend to reside in less infrared-luminous hosts.  The dearth of intrinsically weak AGN 
in the most luminous galaxies may be due to increased dilution by star formation and not necessarily 
caused by a difference in the triggering mechanisms.  
If we missed AGN due to star formation dilution, our AGN census would 
be more incomplete at the IR-luminous end.  However, we already observe a larger AGN fraction 
in more IR-luminous systems, which is opposite to the effects of dilution or dust obscuration. 
Thus, the observed increase in AGN fraction with $L_{IR}$ is robust despite this potential bias.

\subsection{AGN on the SFR$-$M$_{\star}$ Sequence}\label{sec:sSFR}

\begin{figure*}
\epsscale{1.05} \plottwo{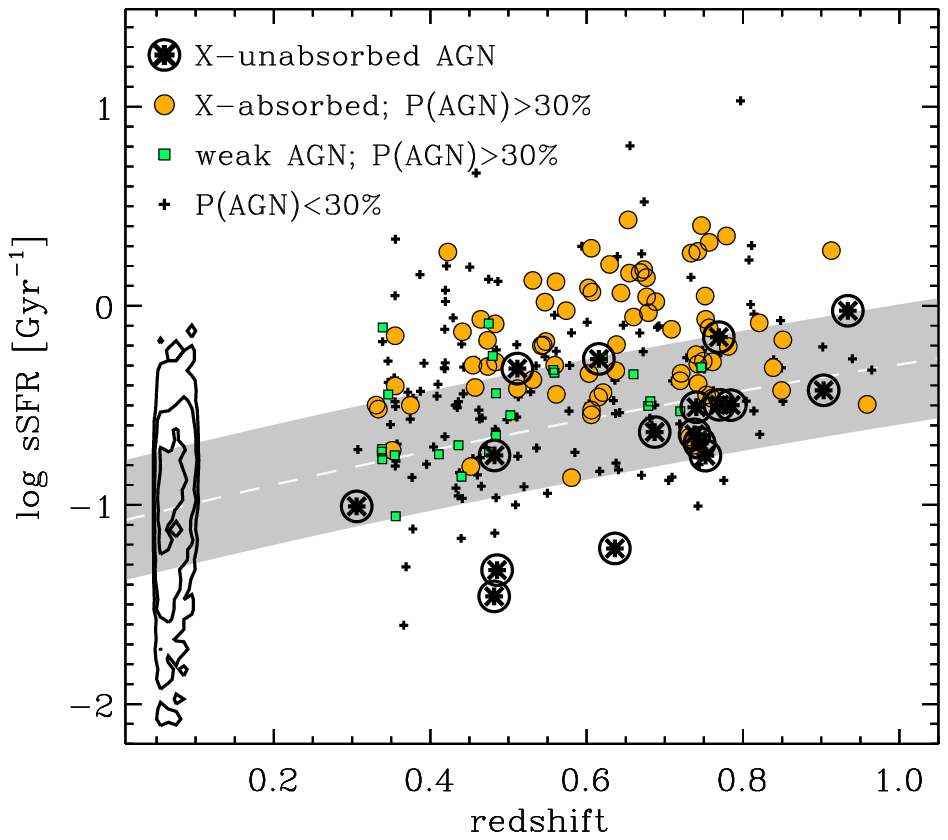}{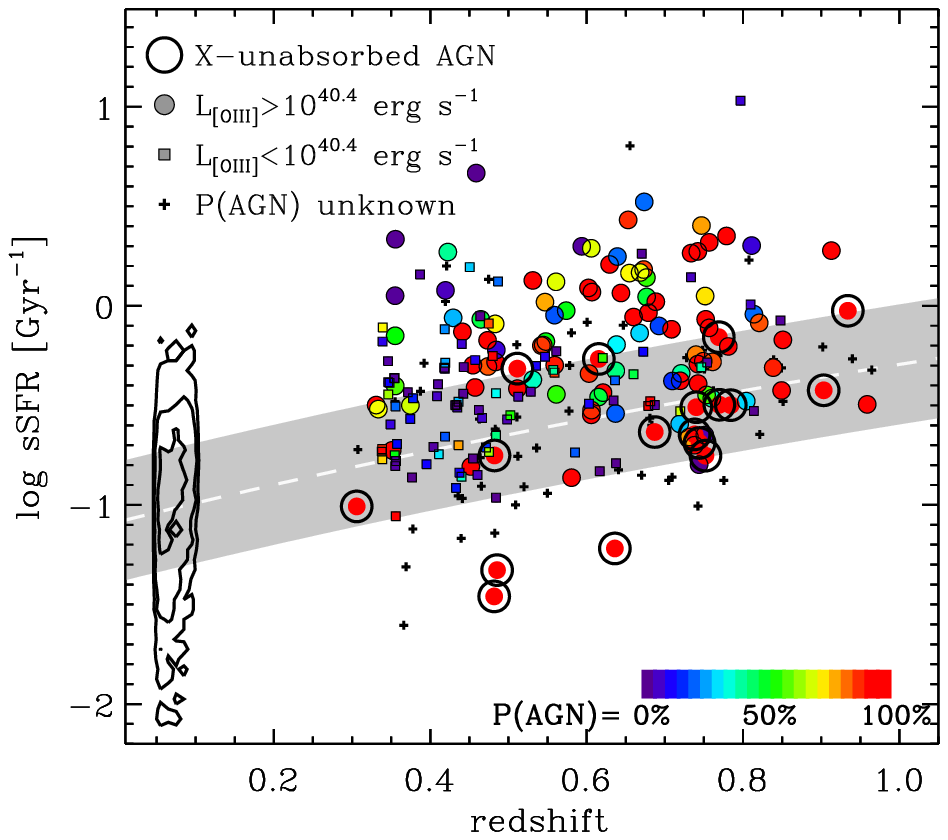}
\caption{
   {\bf Left.}
   Specific star formation rate as a function of redshift.  
   The inter/FIR galaxy sample spans the {\it sSFR sequence}, shown with a gray shaded 
   area, and beyond.  X-ray unabsorbed AGN 
   (black asterisks) mostly lie on or below the sSFR sequence whereas X-ray absorbed AGN (orange circles)
   and intrinsically weak AGN (green squares) tend to have higher sSFRs.  The latter are shown only for 
   objects with an AGN probability percentage $>30$\% but the right hand panel conveys the information on 
   the individual P(AGN) percentages.  
   {\bf Right.}
   Same as left panel, except that the symbols are color-coded according to P(AGN) when available 
   (otherwise shown with black plus symbols).  
   The shape of symbols indicate the \oiii\ luminosity (filled circles if $>10^{40.4}$\,erg\,s$^{-1}$; 
   filled squares otherwise). X-ray unabsorbed AGNs are furthermore marked with large open circles.
   Contours indicate the distribution of the Low/FIR sample on the sSFR-$z$ plane.
   }\label{fig:AGN_sSFR}
\end{figure*}

The high incidence of AGN in infrared-luminous hosts may hint at a link between black hole 
growth and SFR.  The SFR is known to evolve with redshift and to correlate strongly 
with stellar mass at a given redshift.  The correlation between the SFR and stellar mass has 
been reported in several studies \citep[e.g.,][]{bri04,noe07,elb07}. 
Assuming a linear relation between SFR and stellar mass at a given redshift\footnote{The index $\alpha$ 
of the relation SFR $\propto$ M$_{\star}^\alpha$ is consistent with unity within the uncertainties 
and within the spread of values published in the literature but the exact value and whether it changes with 
redshift remains under debate.}, 
the evolution of the normalization has been parameterized by \citet{elb11} as:
\begin{equation}
sSFR_{sequence}(z)~{\rm [Gyr^{-1}]} = 26\times t_{cosmic}^{-2.2} 
\end{equation} 
where $t_{cosmic}$ is the cosmic time in Gyr at the redshift $z$ of interest.  
Here, we study the specific star formation rate, sSFR, of the AGN hosts relative to the remainder 
of the inter/FIR sample. 
The X-ray unabsorbed AGNs tend to reside in galaxies on or slightly below the sSFR 
sequence, in agreement with results from \citet{mul11b}), while the X-ray absorbed 
and/or X-ray weak AGNs are hosted by galaxies with higher sSFRs (Figure~\ref{fig:AGN_sSFR}).

%---------------------------------------------
% Figure 7 here
%---------------------------------------------
%---------------------------------------------

\begin{figure*}
\epsscale{1.05} \plotone{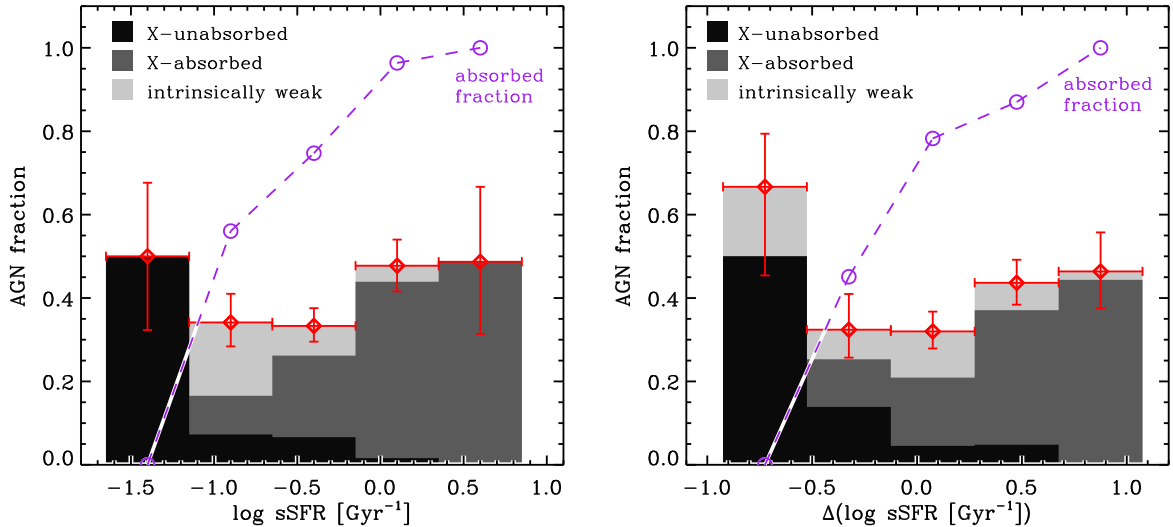}
\caption{
   {\bf Left.} 
   Fraction of galaxies hosting active nuclei as a function of their specific star formation rate 
   (red diamonds). 
   {\bf Right.}
   Fraction of galaxies hosting active nuclei as a function of their distance from the specific 
   star formation rate sequence (red diamonds).  
   In both panels, the AGN categories are distinguished with the shades of gray as labeled.     
   The X-ray absorbed fraction among the moderately-luminous AGN population (e.g., excluding the 
   weak AGN category), is shown with the open circles and dashed purple line.  
   Horizontal error bars show the bin width and vertical error bars show the 68.3\% 
   confidence intervals for fractions, determined with Bayesian binomial statistics 
   following the formalism of \citet{cam11}.
   }\label{fig:AGNfrac_sSFR}
\end{figure*}

The overall AGN fraction appears to be independent of both the sSFR and $\Delta$log(sSFR), 
the relative offset from the redshift-dependent sSFR sequence:
\begin{equation}
\Delta\log(sSFR) \equiv \log(sSFR) - \log(sSFR_{sequence}(z)) 
\end{equation}
(Figure~\ref{fig:AGNfrac_sSFR}).
Contrary to the near constancy of the AGN fraction with respect to the sSFR, we find a strongly increasing 
fraction of X-ray absorbed AGNs among the moderately-luminous AGN population as sSFR (or $\Delta$log(sSFR)) 
increases.  Combined, these results suggest that the AGN obscuration depends strongly 
on the sSFRs of the host galaxies, but the total incidence of AGNs (probed by the AGN fraction) does not.  

%---------------------------------------------
% Figure 8 here
%---------------------------------------------
%---------------------------------------------

\subsection{X-ray Absorbed AGN}\label{sec:absorb}

\begin{figure*}
\epsscale{1.05} \plotone{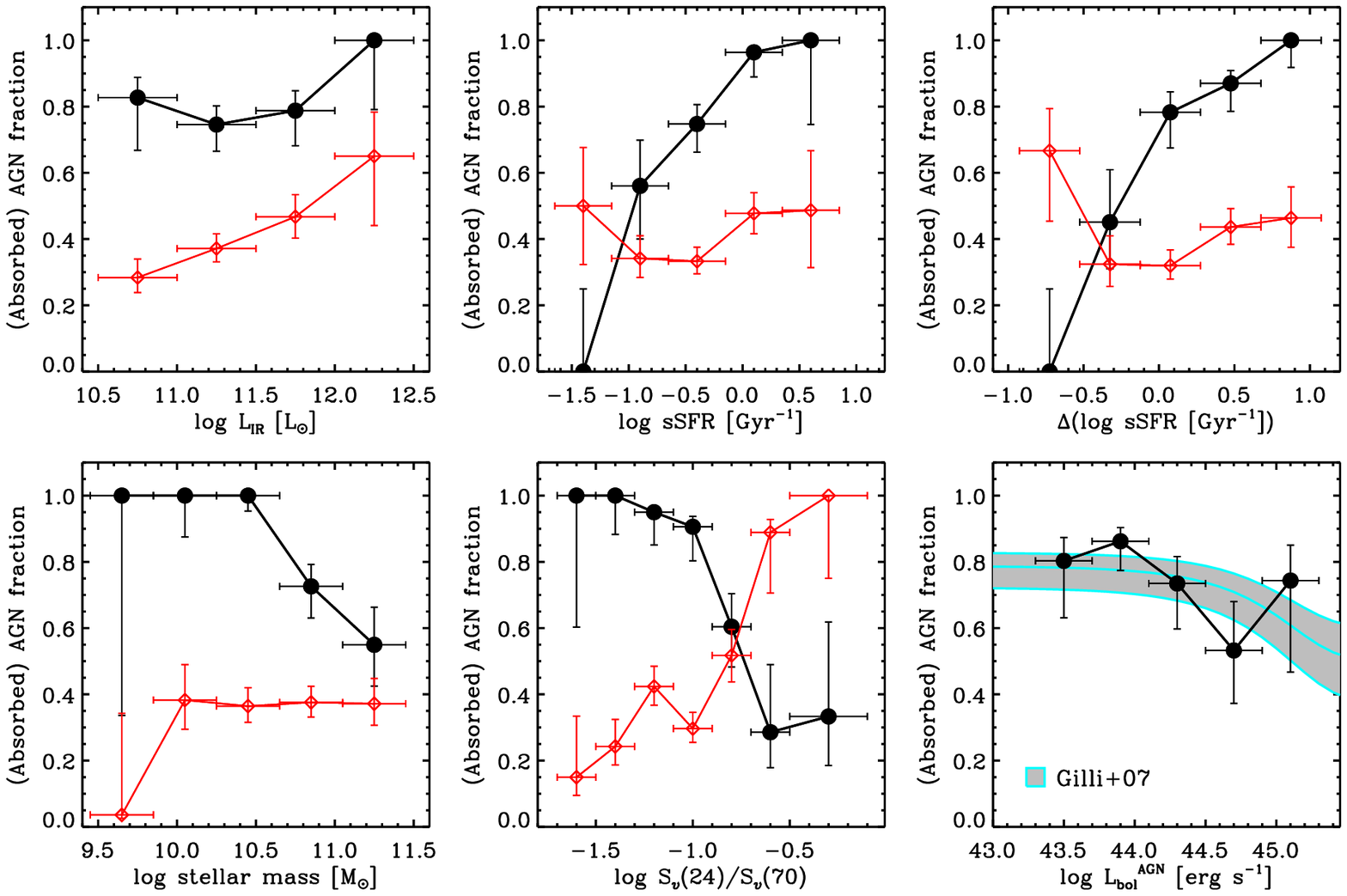}
\caption{
   AGN fraction (red diamonds) and fraction of the AGNs that are X-ray absorbed (filled black circles) as a function of the 
   following parameters: infrared luminosity, specific SFR, distance from the sSFR sequence ($\Delta$sSFR), 
   stellar mass, mid-to-far infrared color, and AGN bolometric luminosity. In the last panel, the X-ray obscured fraction 
   from \citet{gil07} is shown with the shaded area, where we applied the same bolometric correction factor to the X-ray 
   luminosity as in the rest of this work (1.4~dex). 
   Horizontal error bars show the bin width and vertical error bars show the 68.3\% 
   confidence intervals for fractions, determined with Bayesian binomial statistics 
   following the formalism of \citet{cam11}.
   }\label{fig:AbsFrac}
\end{figure*}

As shown in Section~\ref{sec:irAGN} and summarized in Table~\ref{tab:samples}, X-ray 
absorbed AGNs are more common in actively star-forming (FIR-selected) galaxies (24\%) 
than are X-ray unabsorbed AGNs (7\%).  
In what follows, we define the absorbed AGN fraction as the number of X-absorbed AGN over 
the combined number of X-absorbed and X-unabsorbed AGN, i.e., without taking into account 
the intrinsically weak systems.  The behavior of black hole growth and AGN absorption is 
investigated as a function as host galaxies properties. Figure~\ref{fig:AbsFrac} shows that 
the occurrence of AGN increases with infrared luminosity, tracing SFR, and is  
fairly flat with stellar mass down to $10^{10}$\,M$_{\sun}$. The uncertainty in the lowest mass bin 
is too large to constrain the behavior of the AGN fraction at masses $<10^{10}$\,M$_{\sun}$.  
The AGN fraction against the sSFR is consistent with a constant value around $\sim$35-40\% 
on and outside the sSFR sequence.  Lastly, the AGN fraction increases with warmer 24-to-70\,$\mu$m colors 
as expected from an AGN contribution to dust heating (Section~\ref{sec:ir_emline}).

The behaviour of the absorbed AGN fraction differs from that of the occurrence of AGN in galaxies.  
The fraction of AGNs that are absorbed is fairly constant with IR luminosity with perhaps a mild increase at the highest $L_{IR}$, 
while it decreases with increasing stellar mass, and increases sharply with increasing sSFR.
These trends suggest that AGN X-ray absorption may not be as sensitive to the total gas mass (somewhat traced 
by $L_{IR}$) as to the gas fraction and/or gas density, traced by sSFR.  The absorbed fraction decreases with warmer 
24-to-70\,$\mu$m colors, suggesting a smaller relative contribution of star formation in less absorbed AGN.  
Lastly, the absorbed fraction decreases with increasing bolometric luminosity of the AGN, in agreement 
with results reported previously \citep[e.g.][]{ued03,laf05,tru09}.  \citet{gil07} proposed an analytical description 
of the absorbed fraction adjusted to represent the data points of \citet{ued03,laf05} with their model m2, shown in the 
bottom right panel of Figure~\ref{fig:AbsFrac}.  

%---------------------------------------------
% Figure 9 here
%---------------------------------------------
%---------------------------------------------

\subsection{Effect of AGN on Mid-to-Far Infrared Color}\label{sec:ir_emline}

\begin{figure}
\epsscale{1.05} \plotone{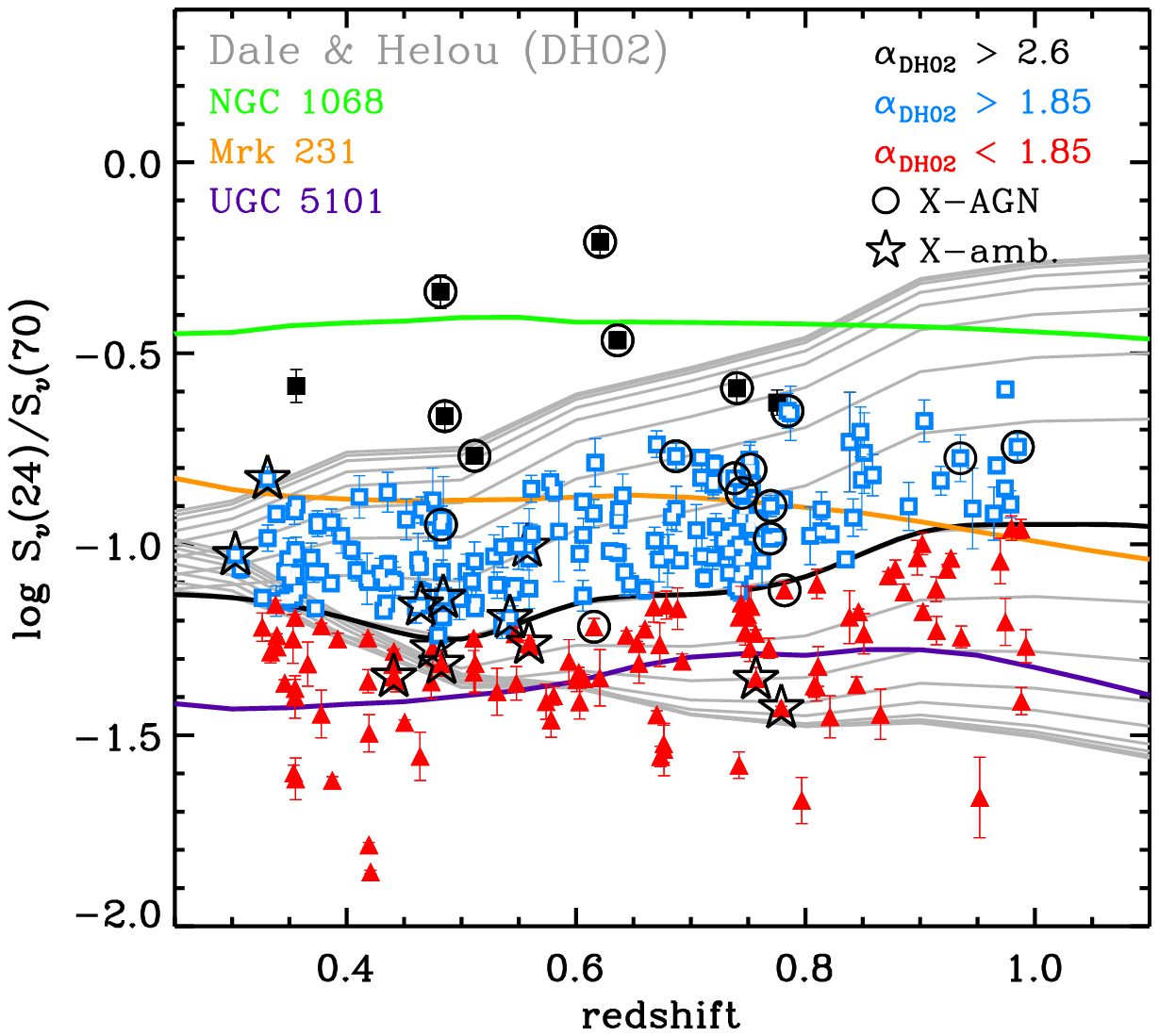}
\caption{Observed 24-to-70\,$\mu$m flux density ratio as a function of redshift for galaxies in 
   GOODS-N and EGS.  The observed values are compared directly to redshifted model templates from 
   \citet{dal02}, applicable to star-forming galaxies (gray lines).  The black line shows the 
   DH02 template chosen to split our sample between high $S_{24}/S_{70}$ (blue squares) and low 
   $S_{24}/S_{70}$ ratios (red triangles), with the most extreme $S_{24}/S_{70}$ highlighted with filled
   black squares.  X-ray AGNs and X-ray ambiguous sources are respectively 
   marked with open circles and star symbols.
   For visual comparison, we include redshifted tracks for three nearby AGNs: NGC~1068 
   (Seyfert 2, in green), Mrk~231 (Seyfert 1, in orange), and UGC~5101 (obscured or buried AGN,
   in purple).
   }\label{fig:S24S70}
\end{figure}

The observed 24-to-70$\mu$m flux ratios of the Inter/FIR galaxy sample 
span a broad range of values (Figure~\ref{fig:S24S70}), extending beyond the range 
of the IR SED templates from \citet[hereafter DH02]{dal02}, 
especially at lower redshift\footnote{The DH02 models were constructed using IRAS 
sources, and so are deficient in galaxies whose FIR is dominated by cold dust such as 
those selected in longer wavebands with, e.g., {\it Herschel} \citep{smi12}.}.  
We use the DH02 template that splits the sample in comparable numbers on either side as a  
redshift-dependent dividing line between galaxies with high and low $S_{24}/S_{70}$ galaxies.  
The chosen template has parameter $\alpha$ = 1.85, where $\alpha$ is the index of the 
power-law relating the mass of dust heated to the intensity of the interstellar radiation
field responsible for the heating, see Equation~1 of DH02.
According to the method described by \citet{marci06}, this SED corresponds to a total 
infrared luminosity of $10^{11}~L_{\sun}$.
X-ray identified AGNs tend to reside above the line and reach the largest values of $S_{24}/S_{70}$.

%---------------------------------------------
% Figure 10.
%---------------------------------------------
%---------------------------------------------

\begin{figure*}
\epsscale{1.15} \plotone{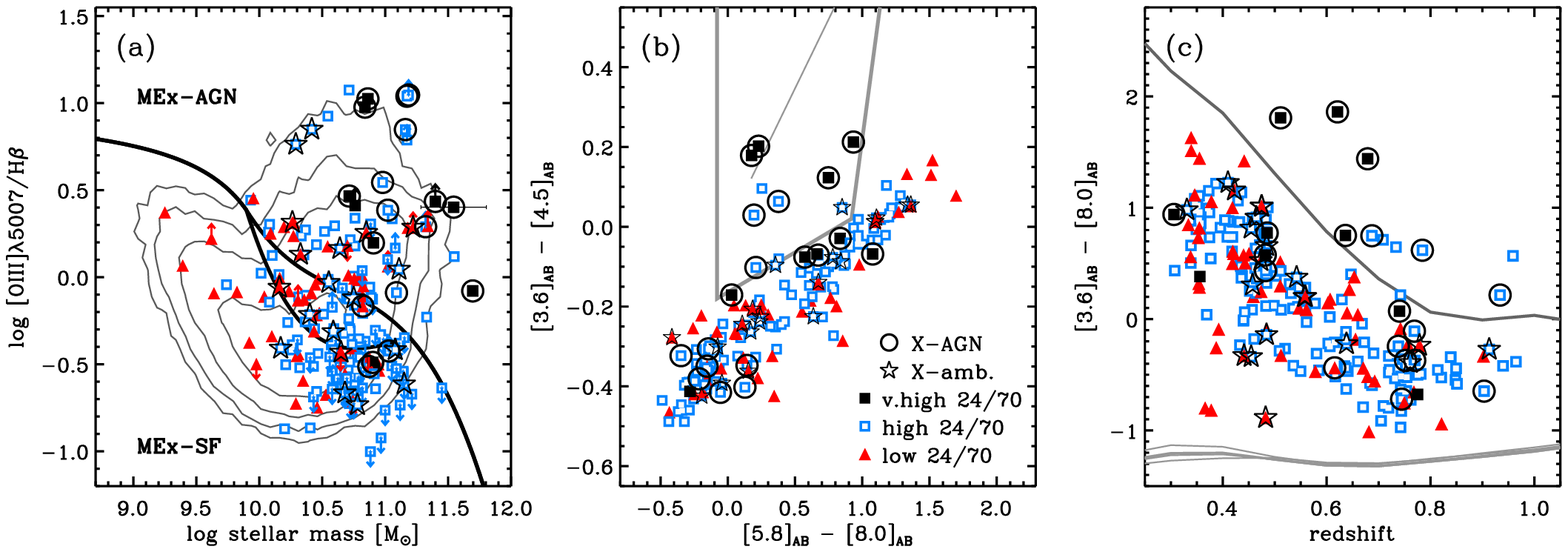}
\caption{
   AGN diagnostic diagrams introduced in Figure~\ref{fig:intromex}.  
   Symbols are keyed to the mid-to-far-infrared color 
   $S_{24}/S_{70}$ (red triangles for low values; open blue squares for high values; 
   filled black squares for very high values) and to the X-ray classification 
   (open circles for X-AGN and open star symbols for X-ambiguous).
   (a) The MEx AGN diagnostic diagram.  The empirical 
   curves (solid lines) split the diagrams into galaxy spectral types 
   as labeled, with a MEx-intermediate region between the MEx-AGN and 
   MEx-SF classes.  Contours show the 
   SDSS low-$z$ sample (evenly spaced logarithmically).
   Objects with low values of $S_{24}/S_{70}$ tend to be centrally 
   distributed on the MEx diagram while the location of the 
   high-$S_{24}/S_{70}$ galaxies extends to both extremes 
   of the \oiiihb\ range.  Galaxies with very high $S_{24}/S_{70}$ 
   are exclusively on the AGN tail of the distribution toward 
   high values of \oiiihb.
   (b) IRAC color-color diagram with the AGN region (thick solid line) defined 
   by \citep{ste05}, and the power locus (thin solid line) by \citep{don07}.  
   Galaxies with the highest \mipscol\ are located within or nearby 
   the boundaries of the AGN region, as expected if there were a 24$\mu$m 
   excess caused by AGN-heated dust emission.
   (c) IRAC color as a function of redshift.  The AGN candidates lie above the 
   dark gray dividing line and span a broad range of $S_{24}/S_{70}$ values although  
   72\% (13/18) have high or very high ratios (open and filled squares).  
   The light gray tracks at the bottom define the lower envelope expected 
   from BC03 stellar population templates with no dust emission.
}\label{fig:agnIRcol}
\end{figure*}

With Figure~\ref{fig:agnIRcol}, we study the mid-to-far infrared color of the AGN hosts 
(including X-ray, IR, and MEx AGNs) with respect to the rest of the star-forming galaxies in our sample.
Galaxies with a low $S_{24}/S_{70}$ (red triangles) occupy predominantly the star-forming 
and intermediate regions of the MEx diagram (Panel a), and the  
star-forming region of the IRAC two-color diagram (Panel b).  
Galaxies with a high $S_{24}/S_{70}$ (blue squares) are distributed across the 
full range of parameter space with the most extreme cases (filled black squares) 
almost exclusively classified as AGN on the MEx diagram, and in or near the AGN region 
on the IRAC two-color diagram.  Galaxy mid-to-far-infrared colors are more evenly 
distributed on the IRAC single-color diagram (Panel c) but there is an excess 
of the most extreme infrared colors (filled black squares) in the AGN region.
Overall, galaxies with the highest $S_{24}/S_{70}$ values are much more likely to be 
identified as AGN with any (or all) of the AGN tracers, especially X-rays and optical lines.
This result indicates that the active nuclei within these galaxies significantly contribute to 
the heating the dust whose emission is measured by the 24 and 70~$\mu$m passbands at these redshifts.

%---------------------------------------------
% Figure 11
%---------------------------------------------
%---------------------------------------------

The AGN-heated contribution to dust emission is expected to peak at shorter wavelengths ($\sim10-20~\mu$m) 
compared to dust heated by star formation processes, and hence produce an enhanced mid- to far-IR ratio.  
The average AGN SED template of \citet{elv94} suggests this feature \citep[see also][for the intrinsic AGN 
IR SED of more moderate luminosity AGN]{mul11a}.  
%% Siebenmorgen and other models
\citet{vei02} found that the spectra of nearby ULIRGs with 
warmer IRAS 25-to-60~$\mu$m color ($S_{25}/S_{60} > 0.2$) are quasar-like whereas the spectral features of 
cooler ULIRGs are similar to LINER or purely star-forming galaxies.  
Similarly, we observe that the galaxies with the highest \mipscol\ values were selected as 
X-ray AGN (Figure~\ref{fig:S24S70}), MEx-AGN (Figure~\ref{fig:agnIRcol}a) and mostly also as 
IRAC AGN (Figure~\ref{fig:agnIRcol}b, c).
The expectations appear to hold well for these few extreme systems, but whether there is also a trend for the 
more normal, less extreme AGNs is less clear.

\begin{figure}
\epsscale{1.05} \plotone{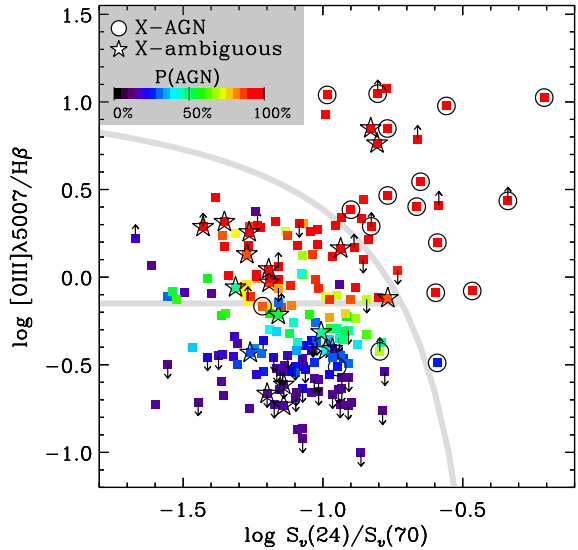}
\caption{
   Combined optical/IR diagram showing the \oiiilam/\hb\ emission-line 
   flux ratio as a function of a mid- to far-IR color (24-to-70$\mu$m).  
   The 70$\mu$m galaxy sample was classified based on the MEx diagram 
   (color keyed to P(AGN) as shown on color bar) as well
   as from their X-ray properties (star symbols for X-ray 
   starbursts/ambiguous systems, and black circles for
   X-ray AGNs).  The lines represent an empirical division 
   between the most extreme AGNs (top right), the more common AGNs (middle) 
   and star-forming galaxies (bottom left).
   }\label{fig:diagDTEx}
\end{figure}

Figure~\ref{fig:diagDTEx} presents another diagnostic by combining the \oiiilam/\hb\ and $S_{24}/S_{70}$ 
ratios.  There are 202 galaxies in our sample with all required measurements, 
including galaxies with an upper limit for either \oiiilam\ or \hb\ (but not both).  
The bulk of the sample tends to occupy the bottom left part of the plot with several outliers toward 
the upper right, which are the more extreme and easily identified AGN (according to at least one of 
the X-ray and the MEx classification schemes).  

We define an empirical curve to divide the region dominated by X-ray AGNs from the rest of the sample:
\begin{equation}\label{eq:OiiiHb_2470}
{\rm log}(\oiii/\hb) = \frac{0.6}{{\rm log}(S_{24}/S_{70}) + 0.28} + 1.2
\end{equation}
A dividing line at fixed \oiii/\hb\ ratio (log(\oiii/\hb)= -0.15) separates 
the optically selected AGNs (green to red colored symbols) from the star-forming galaxies
(purple to blue colored symbols).  Given the fact that the galaxies between the lines are not 
detected in hard X-rays and given their range of \oiii/\hb\ values, they may be composite SF/AGN galaxies 
and possibly LINERs \citep[though we expect very few LINERs given the FIR selection; see, e.g.,][]{yua10}. 
Interestingly, the X-ray undetected AGN candidates and star-forming galaxies appear to have a 
very similar distribution in their 24-to-70$\mu$m colors.  
These systems may have a weak AGN whose contribution to the 
dust emission is small, or a more luminous AGN concurrent with an elevated SFR and thus also 
resulting in a small \emph{relative} contribution.  Alternatively, the IR SED shape might have a more 
complicated dependence on AGN heating owing to the geometry of the system. 
More generally, this result implies that factors other than the 
presence of AGN should be taken into account in order to explain the spread in $S_{24}/S_{70}$
values.

This analysis is analogous to that presented by \citet{kew01a}, where the authors derived a mixing
line going from low values of IRAS 25/60~$\mu$m flux ratio and \oiii/\hb\ ratio toward higher values 
as the AGN fraction increases.  Furthermore, they found AGN fractions of 73\% and 77\% in galaxies with 
$\log$($S_{25}/S_{60}$) $>$ -0.6 and -0.5, respectively.  
Although probing somewhat different rest-frame wavelengths, our results are consistent with 
the same general picture, as all galaxies with $\log$(\mipscol) $>$ -0.6 satisfy our AGN selection criteria.
The results of increasing incidence of AGN in systems with warmer mid-to-far infrared colors 
from \citet{kew01a} and \citet{vei02} were derived for nearby galaxies.  Figure~\ref{fig:diagDTEx} is the 
higher-redshift analog.  

%---------------------------------------------
% Figure 12
%---------------------------------------------
%---------------------------------------------

\begin{figure}
\epsscale{1.05} \plotone{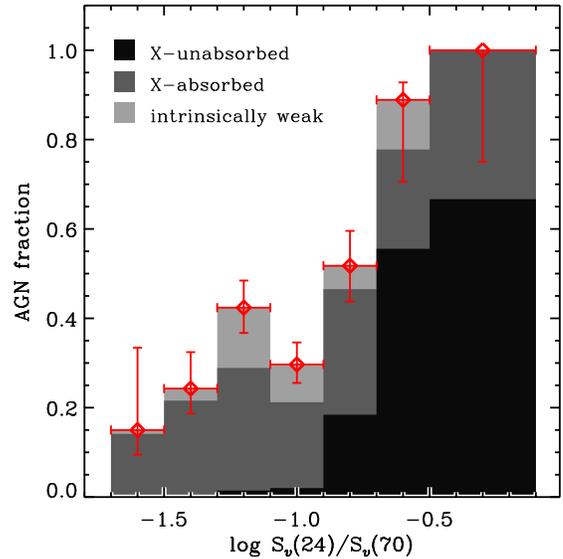}
\caption{
   AGN fraction as a function of mid- to far-IR color (24-to-70$\mu$m). 
   The FIDEL 70$\mu$m galaxy sample was classified as AGN based on all of the AGN diagnostics 
   used in this paper: X-rays, MEx diagram, and IRAC colors.  The split into AGN sub-categories 
   is described in Section~\ref{sec:irAGN} and is identical to the right panel of Figure~\ref{fig:AGNfrac_IR}.
   Horizontal error bars show the bin width and vertical error bars show the 68.3\% 
   confidence intervals for fractions, determined with Bayesian binomial statistics 
   following the formalism of \citet{cam11}.
      }\label{fig:fracIRcol}
\end{figure}

The overall trend between the presence of AGN and the mid-to-far-IR color \mipscol\ is 
shown in Figure~\ref{fig:fracIRcol}, where the AGN fractions are calculated using the 
same methodology as in Section~\ref{sec:irAGN}.  
We find an increasing AGN fraction with warmer \mipscol\ values.
While the warm color side appears to be driven by X-ray detected AGNs,  
removing these objects from the sample reduces the linear Pearson coefficient only from 0.94 to 0.88.
Overall, we thus recover the expected trend from dust heating by AGN in the sample 
of FIR-selected galaxies used in this work.

%---------------------------------------------
% Figure 13
%---------------------------------------------
%---------------------------------------------

%% SECTION 6

%----------------------------------------------------------------------------
\section{Discussion}\label{discu}

\subsection{Caveats}\label{sec:caveats}

There are three main caveats in this analysis although none of them is expected to significantly alter our results.  
The first is that total infrared luminosities, which we interpret as measures of SFR, 
contain mixed contributions from dust heated by both stars and AGN.  While longer wavelengths ($70\mu$m) 
are less subject to AGN contamination than, say, 24$\mu$m, they can still have non-zero contribution from 
AGN-heated dust emission.  
Recent {\it Herschel} studies have shown that the FIR colors of galaxies with an AGN do not 
differ from those of purely star-forming galaxies \citep{hat10,elb10}, indicating that on average 
AGN contamination does not dominate at $\lambda$(observed)$\geq 70~\mu$m.  
We mitigate against this effect by not using the shorter wavelength data ($\lambda \leq 24~\mu$m) 
when calculating $L_{IR}$ for the FIR-selected sample (Section~\ref{sec:sfr}).

The second caveat concerns the overall completeness of the AGN selection methods used in this work.  
As mentioned in Section~\ref{sec:irAGN}, some objects have an unknown classification because of the low 
S/N of their emission lines.  As a consequence, the AGN fraction that we derive for the overall 
FIR-selected sample (i.e., with and without emission line detections) may be underestimated as 
we consider these unclassified galaxies to be purely star-forming unless they fulfill the X-ray or IR AGN 
selection criteria.  Taking into account that some of 
the unknown class are AGN would strengthen our conclusion that AGN are ubiquitous among FIR-selected 
galaxies given that we already find a large AGN fraction.  We computed a maximum AGN fraction 
assuming that all of the unknown class galaxies host an AGN: it would rise from 37\% to 62\%, but this 
scenario is unlikely given that the unknown galaxies do not fulfill the emission line 
criteria and that emission-line galaxies were shown to have a higher incidence of AGN than 
their non emission-line counterparts \citep[e.g.][]{yan06}.
We test this trend with our own sample by comparing the X-ray and IRAC AGN fractions for the 
global FIR sample and the emission-line FIR sample.  These methods suggest a $2\times$ higher AGN
fraction in emission-line galaxies.  
There are 12$^{+4}_{-2}$\% (18/145) X-AGN and IRAC-AGN in the emission-line subsample, but only 6$\pm$2\% 
(7/125) in the remaining subsample, resulting in a global fraction of 9\% (25/270) 
when considering all FIR-selected galaxies regardless of the detection of \oiii\ and \hb\ emission lines. 
Among objects with AGN classifications, it is also possible that the MEx diagnostic diagram misses 
AGNs preferentially in low-mass galaxies, especially if there is also ongoing star formation 
diluting AGN emission signatures.  Such AGNs are fairly rare in the nearby universe 
but would not be accounted for if they were more common at higher redshift.

The third caveat concerns the use of \oiiilam\ as an indicator of bolometric AGN luminosity.  
On the one hand, \oiii\ emission could include a contribution from star formation yielding 
an overestimate of the total AGN power.  This is a more serious concern for low-metallicity 
galaxies. \citet{kau03c} found that in high-metallicity galaxies, the \oiiilam\ line is the 
least contaminated of the strong emission lines measured in optical spectra, with a flux fraction 
of 7\% from star formation and the remaining 93\% from AGN induced emission.
The stellar mass-metallicity ($M_{\star}-Z$) relation \citep{tre04} furthermore implies that 
metal poor galaxies have small stellar masses. 
Given that all the FIR-selected galaxies are fairly massive, they likely have a fairly elevated 
metallicity and their \oiii\ lines are likely dominated by AGN emission.  
However, the mass-metallicity relation has been shown to evolve with redshift 
\citep{sav05} and/or SFRs of the galaxies \citep{man10} in the sense that the galaxies in the current sample 
could have a slightly lower metallicity than galaxies of like stellar masses in the nearby universe. 
We thus compared \Loiii\ to $L_{IR}$, tracing the SFR, in order to check whether the MEx-AGNs present 
\oiii\ excess due to their elevated SFRs rather than true AGN excitation.  
Both X-ray absorbed and unabsorbed AGNs show the same \oiii-to-IR excess with respect to the subsamples 
of Weak AGNs and star-forming galaxies (Figure~\ref{fig:distr}). 

On the other hand, \oiiilam\ emission may be affected by dust obscuration.  Even though we use \oiii\ as a more isotropic tracer than 2$-$10~keV, it is not in fact perfectly isotropic \citep[e.g.][]{dia09}.  Dust obscuration would have the 
opposite effect, and would lead to an underestimate of the intrinsic AGN luminosity when not 
applying a correction for extinction.  We conservatively choose not to correct for dust 
obscuration, which means that the X-ray absorption may be underestimated and similarly, 
the number of absorbed AGN may be a lower limit.  Given that we already 
find a high incidence of absorbed AGN among FIR-selected galaxies, our result would only be 
strengthened if we indeed underestimated the number of absorbed AGNs.

\subsection{Link between AGN Obscuration and Host Galaxies}\label{sec:linkHost}

A high absorbed fraction among AGNs at high redshift may be expected because in addition 
to small-scale (torus) absorption, there were important gas reservoirs in galaxies at $z>1$ 
\citep{dad10,tac10} that can potentially contribute to absorbing X-rays. Models of gas-rich 
galaxies predict high column densities (reaching the Compton-thick regime with $N_H>10^{24}$\,cm$^{-2}$) 
along several lines-of-sight in both isolated gas-rich unstable disks \citep{bou11} and in major 
mergers \citep[e.g.,][]{hop06}. 
Furthermore, analyses of the cosmic X-ray background infer an important population of 
X-ray absorbed AGNs at $z<1$ \citep{com95,mer04}.
Therefore, a high fraction of hidden AGNs should be expected at increasing redshift.  

Is the absorption occurring on small scales or galaxy scales or both?  In the former 
case, one would expect that AGN absorption would not correlate with galaxy-scale 
properties\footnote{
Although recent observations suggest some (at least small degree) alignment of AGN and 
their host galaxies \citep{lag11}.}.
Interestingly, we found that the fraction of the AGNs that are absorbed increases sharply with 
the sSFRs (Figure~\ref{fig:AbsFrac}).  Before discussing the physical implications, 
we further examine X-ray absorption in relation to the infrared luminosities of the host 
galaxies with an extension to fainter host galaxy samples. 

We start from AGNs that have both X-ray and \oiii\ detections in the Inter/All sample 
(dubbed the X$+$\oiii\ sample), and 
divide this population according to their detections in the MIR only (FIDEL 24~$\mu$m) 
or in the FIR (FIDEL 70~$\mu$m).  The 24$\mu$m comparison sample 
has a much fainter ($\sim10\times$) $L_{IR}$ sensitivity limit than the 70~$\mu$m 
sample (Figure~\ref{fig:Lir}).  The remainder of the X-ray+\oiii\ AGN sample 
lacks FIDEL 24~$\mu$m detections and thus probes yet lower IR luminosities.  
These three subsamples have significant differences in sensitivity in 
terms of the SFR of their hosts but we otherwise constrain this portion of the analysis to 
galaxies that have $\log(M_{\star})>10^{10}$~M$_{\sun}$ and P(AGN)$>$30\%\footnote{We set P(AGN)$=1$ 
if the source is selected as an X-ray AGN.}. 
With comparable masses and fairly secure AGNs, the 70~$\mu$m subsample will include 
galaxies with the highest sSFRs whereas the 24~$\mu$m and X+\oiii\ subsamples probe 
down to lower sSFRs given a range of masses and redshifts.  

\begin{figure}
\epsscale{1.0} \plotone{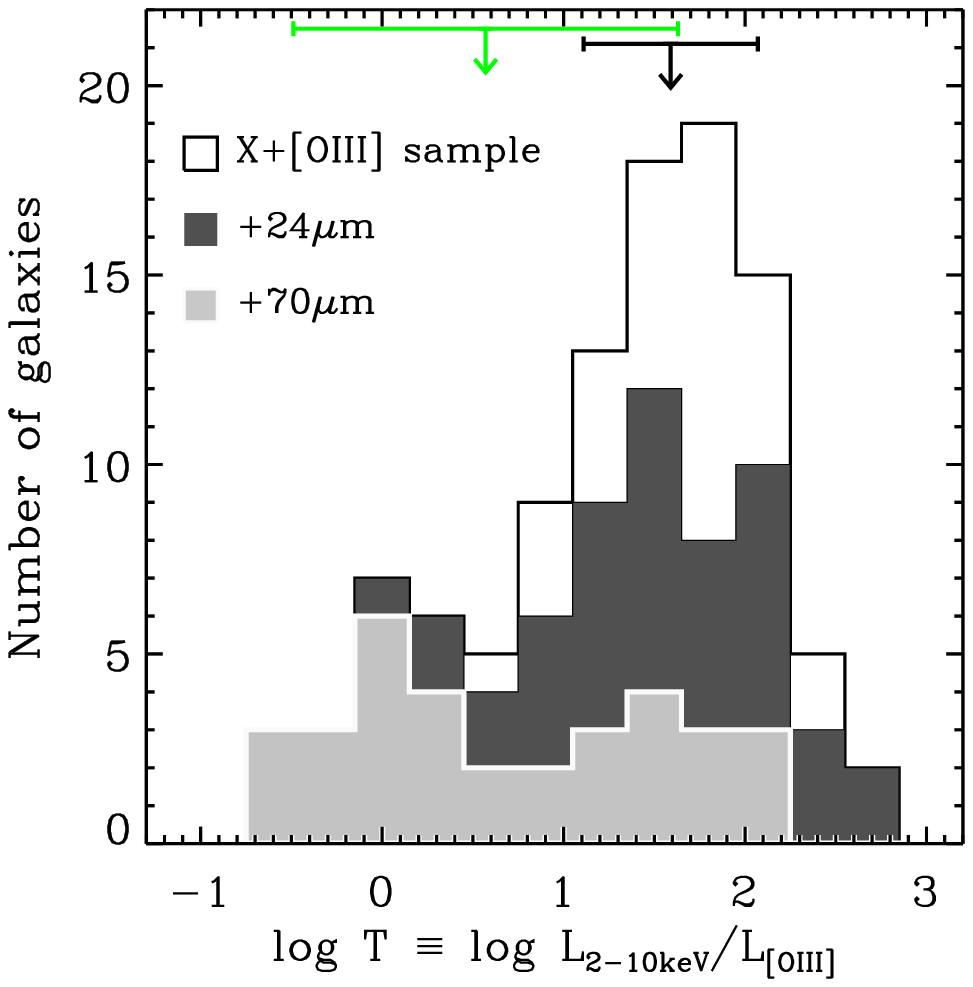}
\caption{
   Distribution of the Compton-thickness parameter $T \equiv L_{2-10keV}/L_{\oiiilam}$ for an 
   X-ray$+$\oiii-selected sample of AGNs (black line).  The distributions for the subsamples of 
   galaxies detected at 24~$\mu$m and at 70~$\mu$m are overplotted in dark and light gray, respectively.  
   The most absorbed systems (with lower values of $T$) are preferentially detected at 70$\mu$m.  
   This subsample displays a much broader distribution of thickness parameter and a larger relative 
   number of absorbed AGN compared to both the 24$\mu$m sample and the parent X-ray$+$\oiii-selected sample.  
   Arrows at the top of the figure indicate the mean and standard deviation for nearby AGNs that are 
   either obscured, Type~2 (0.59$\pm$1.06\,dex; in green) or unobscured, Type~1 
   \citep[1.59$\pm$0.48\,dex; in black;][]{hec05}. 
   }\label{fig:OiiiX_IR}
\end{figure}

The distributions of the Compton-thickness parameter  ($\log T \equiv \log(L_{2-10keV}/L_{\oiii})$) 
between the FIR (70\,$\mu$m), MIR (24\,$\mu$m) and X$+$\oiii\ samples are shown in Figure~\ref{fig:OiiiX_IR}.  
The FIR-selected subsample tends to host more absorbed AGNs than either of the comparison samples. 
If the parent population of the FIR (light gray) and X$+$\oiii\ (black line without the light gray area) 
subsamples were the same, the probability of drawing both distributions would be 0.0012\% according to a KS test.
It remains very low (0.038\%) for the distributions of the FIR- and MIR-selected subsamples.  
The distribution of the latter is intermediate between the X$+$\oiii\ subsample and the most
IR-luminous (detected at 70~$\mu$m) sample.  Clearly, the majority of the most absorbed systems 
(with Compton-thickness $\log(T)<0.3$; Appendix~\ref{app:Xabsor}) tend to be infrared-luminous 
given their high detection rate at 70~$\mu$m.

%---------------------------------------------
% Figure 14
%---------------------------------------------
%---------------------------------------------

The overall trend is more frequent AGN X-ray absorption in galaxies with higher sSFRs. 
In order for this trend to be caused by geometry of the putative torus from AGN unified models, 
the small-scale obscuring torus would have to be physically and/or optically thicker in FIR-luminous 
galaxies with elevated sSFR.  
Another possibility is that (part of) the gas reservoir available to sustain the elevated SFR 
also acts as an important absorber for the central AGN.  This situation could be increasingly 
common at higher redshifts as the AGN hosts show an increase in SFR \citep{mul10,sha10,mul11b}.  
Our findings imply that it is not just the total SFR (or $L_{IR}$) that is relevant but rather 
the sSFR.  This points to another factor explaining the apparent connection between AGN 
X-ray absorption and host galaxies.  We postulate that it could be in part because the sSFR can 
trace the gas fraction across the sSFR sequence 
\citep[but no longer when reaching the more extreme sSFRs;][]{mag12}, 
and because for the more extreme objects above the main sequence, the enhanced sSFR 
could reflect an increased gas density.  The latter can be caused by major galaxy mergers \citep[e.g.,][]{nara08b,jun09}, 
which are common for high-redshift IR-selected galaxies above the sSFR sequence \citep{kar12} and which 
are postulated to result in an increased star formation efficiency \citep{dad10} due to the 
``compacting'' of the star-forming gas into a dense central starburst.
Recent observations showed a correlation between AGN X-ray absorption and $\sim100$~pc scale 
starburst in nearby Seyferts \citep{dia11}, which can be broadly consistent with either of the above 
scenarios (torus-host galaxy link, or obscuration by $\sim100$~pc scale gas).  Some simulations predict 
a multi-scale connection where the black hole fueling \citep{hop10} and/or torus properties 
\citep{hop12} are related to the nuclear SFR via gas inflow albeit in a complicated and perhaps 
stochastic way. These possibilities are both very interesting as they differ from 
the typical AGN unified models where the very central region is assumed to be somewhat decoupled 
from its large-scale surroundings.

Complementary findings suggest that host-scale obscuration may hinder the detection of optical lines 
in hard X-ray selected but optically-dull AGNs \citep[e.g.,][]{rig06}.  These tend to reside in
 disky edge-on 
hosts and, if all optical signatures are lost, would be missed with the MEx diagnostic diagram.  The host 
obscuration that we find here must have a smaller covering fraction of the narrow line regions in order to 
detect optical nebular line signatures in many cases.

\subsection{Properties of Absorbed and X-unabsorbed AGNs: An Evolutionary Sequence?}\label{sec:prop}

The stellar mass distributions of the weak and absorbed 
AGN samples are similar to one another while the mass distribution of X-ray AGN is skewed 
toward higher stellar masses.  According to Figure~\ref{fig:agn}(a), the secure X-ray AGN all have a high 
stellar mass $M_{\star}>10^{10.6}~M_{\sun}$, and seem to be more massive on average than the 
global AGN population.  This suggests that X-ray selected AGN will tend 
to be more easily identified in more massive hosts. However, the statistics are poor with 
only 12 X-ray AGN on the MEx diagram.  Using a larger sample, \citet{mul11b} 
found that X-ray selected AGNs always lie in high stellar mass hosts ($\langle M_{\star} \rangle = 10^{10.7-10.8}~M_{\sun}$) 
across a broad range of redshift (up to $z\sim3$).  This may be due to the presence of more massive black 
holes in these systems although the authors only found a weak correlation between stellar 
mass and X-ray luminosity over the $10^{42}-10^{44}~{\rm erg~s^{-1}}$ range ($M_{\star} \propto L_X^{(1/7)}$).
\citet{air12} explain this feature with observational bias such that more massive 
black holes are observable at a broader range of Eddington ratios than less massive ones, which 
presumably reside in less massive hosts.  While this selection bias will be especially true for 
X-ray selection, it is less severe with optical AGN diagnostics, which are sensitive down to much lower 
accretion rates.  This may explain why the MEx method uncovers AGN down to lower stellar masses 
than the less sensitive X-ray method.  However, every AGN selection is 
incomplete.  Thus, it is of great interest to combine multi-wavelength indicators for the sake of 
completeness.

Relative to X-unabsorbed AGNs, the absorbed AGNs have similar bolometric luminosities but somewhat smaller 
stellar masses (and possibly smaller mass black holes). The absorbed AGN hosts are thus growing both 
their BHs and their stellar content {\it faster} (as they tend to have higher
sSFR), in agreement with the results from \citet{tan12b}.  This is unlikely to result from an artefact 
due to \oiii\ being strongly contaminated by star formation because the \oiii-to-IR 
ratio is elevated in a similar fashion in both the absorbed and X-unabsorbed AGN hosts 
(right panel of Figure~\ref{fig:distr}).

Those systematic differences between the absorbed and unabsorbed AGN hosts may correspond to an evolutionary 
sequence where absorbed AGNs tend to occur at the beginning of a growth cycle with higher specific growth rate and 
elusive X-ray signatures due to gaseous surroundings (on possibly a range of physical scales).  X-ray unabsorbed 
AGNs would then correspond to more mature systems (consistent with higher stellar masses of the hosts) after 
some of the gas responsible for fueling and/or obscuring the central region has been consumed, pushed aside or 
removed.  This idea is similar to the well-known major merger scenario 
\citep[described in more details in the next section]{san88}, but could be more generic and also apply to AGNs 
with intermediate luminosities in isolated but initially gas-rich galaxies, assuming that even in those isolated 
systems, the obscuring material could be gradually removed by the AGN and/or star formation.  
The less extreme AGN luminosities compared to the quasar regime mean that major galaxy mergers may 
not be required in all cases.  

\subsection{Physical Origin of High AGN Fractions}\label{sec:origin}

This study finds a larger AGN fraction (37\%) among IR-luminous, star-forming galaxies 
than previous studies. 
Combining multiwavelength diagnostics from X-ray to radio, \citet{kar10a} found that the AGN
fraction rises steeply with IR luminosity and/or redshift (see the filled black circles 
on the left panel of Figure~\ref{fig:AGNfrac_IR}), but their values are lower than ours 
with an overall fraction $\sim10-20$\% over the redshift and luminosity ranges of interest 
in this paper.  Similarly, \citet{sym10} studied the occurrence of AGN in 70$\mu$m galaxies 
over a similar redshift range and found an AGN fraction $\sim13$\%.  
The main difference between these previous investigations and the current work is the use 
of the MEx diagnostic diagram, which unveils the majority of the AGN including absorbed and 
weak systems.  Some absorbed AGN were successfully selected by previous authors 
with, e.g., mid-IR color diagnostics, but the latter are not as sensitive 
to the presence of active nuclei as the MEx method and are therefore less complete.

What are the physical causes of such a frequent occurrence of AGN?  Is there a connection 
between AGN triggering and the elevated SFR in their host galaxies?  Or between AGN 
absorption and their host galaxies?  The results from this work imply that active 
SMBH growth occurs in parallel with active star formation.  The elevated SFRs are probed 
by the FIR emission and there is a connection between the infrared luminosity and the 
occurrence of AGN.  One interpretation that fits all of these observational 
trends is the major merger scenario \citep[e.g.,][]{san88}.  Major mergers of gas-rich galaxies are 
expected to fuel both starbursts and AGN, with the AGN originally deeply buried in large 
amounts of gas before emerging as X-ray detected AGN.  Qualitatively, our observations
fit a merger scenario with simultaneous star formation and AGN activity.   
In contrast, several authors have pointed out that the morphologies of X-ray AGNs do not 
appear to be mergers more often than non-X-ray AGNs \citep{cis11,sch11,mul11b,koc11}.  
However, these studies did not include the X-ray absorbed systems identified by the MEx method:
it is possible that the X-ray selected AGNs are not associated with mergers (or that they represent
a different evolutionary stage) while the more absorbed AGNs are more directly triggered by mergers.  

An interesting feature from this work is that the rise of the global AGN fraction with $L_{IR}$ 
behaves similarly to the rise in the fraction of galaxies undergoing mergers and interactions 
and/or having an irregular morphology at $z\sim1$ \citep[dashed and dotted lines on Figure~\ref{fig:AGNfrac_IR}, taken from][]{kar12}\footnote{These fractions were calculated 
in a slightly different redshift range ($0.8<z<1.2$ rather than $0.3<z<1$) but are similar to what is expected at 
lower redshifts \citep{kar10b} and so should not be significantly different at the exact redshift range of interest here.}.
In contrast, the fraction of X-ray unabsorbed AGNs is much lower and does not track well with the fraction of mergers 
and/or irregular galaxies.  While somewhat indirect, this argument suggests that X-ray selection of AGN 
may miss the connection between AGN occurrence and galaxy interactions because of a possible mismatch 
between the timescale of AGN X-ray observability of the visibility timescale of mergers.
Future studies of the connection between AGNs and their host galaxies should include 
both absorbed and unabsorbed AGNs to provide us with a more conclusive test.

Despite the apparent consistency of a merger picture, major galaxy mergers may not be 
required to explain our results.  For instance, an alternative way to obtain a high incidence of 
AGNs and also a high incidence of AGN absorption could be large-scale disk instabilities 
in gas-rich star-forming galaxies, as exemplified by high-redshift clumpy galaxies 
\citep{elm07}.  These systems are more common at higher redshift ($z\gtrsim2$) and 
may be explained by simulations \citep{age09} in the context of cold stream accretion 
cosmological models \citep[e.g.,][]{dek09}. Clumpy galaxies persist down to intermediate 
redshift in intermediate mass galaxies \citep{elm09}. 
Unstable (or clumpy) disk galaxies are predicted \citep{bou11} and observed \citep{bou12} 
to efficiently fuel AGNs, including a high likelihood of absorption of X-ray signatures.

Irrespectively of the morphology and kinematics of the host galaxies, work by \citet{tan12b} 
on weak AGNs suggests that, like stronger AGNs, they occur predominantly in massive hosts 
($>10^{10}$\,M$_{\sun}$).  Those authors argued that the stellar mass of the host galaxies 
appears to act as a ``switch'' in triggering AGNs, but the underlying mechanism remains unknown 
given that they do not find dependencies on galaxy morphology.  
The weak AGNs from this Paper would also be consistent with a ``switch'' at
$\sim10^{10}$\,M$_{\sun}$ but we do not probe to sufficiently low masses to really characterize that feature.  
Furthermore, the MEx diagram was designed with SDSS priors and so will only be sensitive to the 
stellar mass range overwhich the bulk of low-redshift SDSS AGNs can be identified with the 
traditional BPT diagrams.

While we cannot distinguish between major galaxy mergers, large-scale violent disk 
instabilities or other triggering mechanism without additional observations, 
it is clear that a complete picture of the growth of galaxies and their central 
SMBHs must include a gas-rich phase with concurrent SF and AGN, where the central 
AGN is often absorbed even for moderate-luminosity AGN 
($10^{43.4}<$\Lbol$<10^{45}$~\ergs, i.e., below the quasar regime).

Lastly, while the AGN {\it fraction} of the intermediate-redshift FIR-selected sample is 
similar to that in nearby FIR-selected galaxies, the {\it number} of 
IR-luminous galaxies is much larger at higher redshift.  This implies that the global 
AGN fraction in star-forming galaxies was higher in the past.  
A greater availability of gas at earlier epochs may be responsible for the more common 
occurrence of highly star-forming (thus IR-bright) galaxies.  The high fraction of AGN, even 
if consistent with that at low-redshift, implies that a much larger population of AGN -- and 
especially X-ray absorbed AGN -- in IR-luminous hosts exists at higher redshift.  
We may be starting to uncover the missing AGN population that has been inferred from 
cosmic X-ray background studies \citep{com95,mus00,ale03,bau04,gil07}.

%----------------------------------------------------------------------------
\section{Summary}\label{summ}

In this paper, we investigated the incidence of AGN among star-forming galaxies at 
intermediate redshift ($0.3<z<1$).  The AGN fraction and 
AGN characteristics (X-ray unabsorbed, X-ray absorbed, intrinsically weak) are 
function of host galaxy properties.  AGNs were identified based on four diagnostics: MEx diagram 
\citep{jun11}, X-ray criteria \citep[similar to][]{bau04}, IRAC color-color \citep{ste05} or 
IRAC [3.6]-[8.0] color versus redshift (Section~\ref{sec:irac}).  
The last two are similar but not identical.  Our main findings are: 

\begin{enumerate}
\item[1.] Combining all AGN diagnostics, the global AGN fraction is 37(30)$\pm$3\% in the 70$\mu$m 
   selected galaxy population including (excluding) AGNs less luminous than $L_{X}=10^{42}$\,erg\,s$^{-1}$ 
   ($L_{bol}=10^{43.4}$\,erg\,s$^{-1}$).  
   This AGN fraction is around a factor of two greater than previous results in similar 
   infrared luminosity and redshift ranges.  

\item[2.] The fraction of star-forming galaxies hosting an AGN increases as a function 
  of $L_{IR}$.  The AGN fractions are very similar to those in 
 nearby ($0.05<z<0.1$) FIR-selected galaxies suggesting mild or no evolution since $z=1$.

\item[3.] The differences between the higher fraction presented here relative to previous studies 
   at comparable redshifts may be accounted for by {\it (i)} heavily absorbed AGNs, and {\it (ii)} 
   intrinsically weaker AGNs.  Absorbed and/or weak systems are more difficult to detect but can be 
   identified thanks to the high sensitivity of the MEx diagnostic.

\item[4.] The fraction of galaxies hosting an AGN appears to be independent of the specific star formation 
   rate of the host, and remains elevated both on the sSFR sequence and above.  In contrast, the 
   fraction of AGNs that are X-ray absorbed increases substantially with increasing sSFR, possibly due to 
   an increased gas fraction or gas density of the main/central star-forming regions in the host galaxies.

\item[5.] AGNs with the most X-ray absorption ($N_H>10^{24}$\,cm$^{-2}$; inferred from $\log(L_{X}/L_{\oiii})<0.3$) 
   tend to reside in IR-luminous galaxies (detected at 70\,$\mu$m).  
   Adding less IR-luminous AGN host samples (undetected at 70\,$\mu$m) 
   mostly contributes unabsorbed and mildly absorbed AGNs (Section~\ref{sec:linkHost}).  
   Together with item 4, this result suggests a connection between 
   the host galaxy's gas content and absorption of the central engine, which can be 
   achieved either through absorption of the AGN X-rays by the ISM of the host galaxy or 
   by a physical, multi-scale link between the host galaxy's gas content and the obscuring torus.

\item[6.] AGN radiation is likely responsible for the highest 24-to-70~$\mu$m flux 
   ratios in the most extreme cases, which tend to be X-ray and optically setected AGNs.  
   However, other factors must be at work to explain the spread of the mid-to-far 
   IR color among star-forming galaxies and galaxies with a weak or absorbed AGN.

\end{enumerate}

%-----------------------------------------------------------------------------
\acknowledgments 
The authors acknowledge useful discussions with James Aird and a helpful report 
from the anonymous referee. The authors deeply thank the many members of the GOODS 
and AEGIS teams who obtained, reduced, and cataloged most of the data used in this paper.  

This work is based in part on observations made with the {\it Spitzer} 
Space Telescope, which is operated by the Jet Propulsion Laboratory, 
California Institute of Technology under a contract with NASA,  
and on observations made with the {\it Chandra} X-ray Observatory, operated by the Smithsonian 
Astrophysical Observatory for and on behalf of NASA under contract NAS8-03060. Additionally, some of the data used herein were obtained at the W. M. Keck Observatory, which is operated as a scientific partnership among the California Institute of Technology, the University of California and the National Aeronautics and Space Administration. The Observatory was made possible by the generous financial support of the W. M. Keck Foundation. The Keck Observatory acknowledges the very significant cultural role and reverence that the summit of Mauna Kea has always had within the indigenous Hawaiian community and appreciate the opportunity to conduct observations from this mountain.

Funding for the DEEP2 survey has been provided by NSF grants AST95-09298, AST-0071048, AST-0071198, AST-0507428, and AST-0507483 as well as NASA LTSA grant NNG04GC89G. The analysis pipeline used to reduce the DEIMOS data was developed at UC Berkeley with support from NSF grant AST-0071048.  The TKRS was funded by a grant to WMKO by the National Science Foundation's Small Grant for Exploratory Research program. 

SJ was partially funded by a FQRNT fellowship (Fonds Qu\'eb\'ecois de recherche sur la nature et la technologie, Canada), a Philanthropic Educational Organization (P.E.O.) Scholar Award.  SJ and FB acknowledge support from the EU through grants ERC-StG-257720 and CosmoComp ITN. JRM thanks The Leverhulme Trust.

{\it Facilities:} \facility{Spitzer (MIPS)}, \facility{Spitzer (IRAC)}, \facility{Keck (DEIMOS)}, \facility{HST (ACS)}, \facility{Chandra (ACIS)} \facility{VLA}

%------------------------------------------------------------------------------------
%% BEGINNING OF APPENDIX
%% \clearpage
\appendix
\section{AGN X-ray Absorption}\label{app:Xabsor}

This Paper uses the Compton-thickness parameter as a proxy for AGN X-ray absorption 
(Sections~\ref{sec:agnsummary} and \ref{sec:absorb}).  
While this approach may not yield accurate measurements of the absorbing column density 
($N_H$), it allows us to separate the most heavily absorbed cases from the rest of the AGN population 
(X-ray unabsorbed and mildly absorbed).  We demonstrate this with the analysis of 
nearby AGNs presented by \citet[hereafter B99]{bas99}.  The B99 sample is not statistically 
complete but it spans a broad range of X-ray absorption properties and includes measurements 
of \oiiilam\ and $2-10$\,keV fluxes, iron $K\alpha$ equivalent widths, and $N_H$ 
from fitting the X-ray observations with a photoelectric cutoff model (assuming 
no components from reflection or stellar emission).  In some cases, the authors note 
Compton-thick signatures with $N_H>10^{25}$\,cm$^{-2}$ if there is also absorption of 
X-rays at energies above 10\,keV, or with $N_H>10^{24}$\,cm$^{-2}$ if there is no data 
available beyond 10\,keV to rule out an absorbing column between $10^{24}$\,cm$^{-2}$ 
and $10^{25}$\,cm$^{-2}$ (see their Table~2).

\begin{figure}
\epsscale{0.6}\plotone{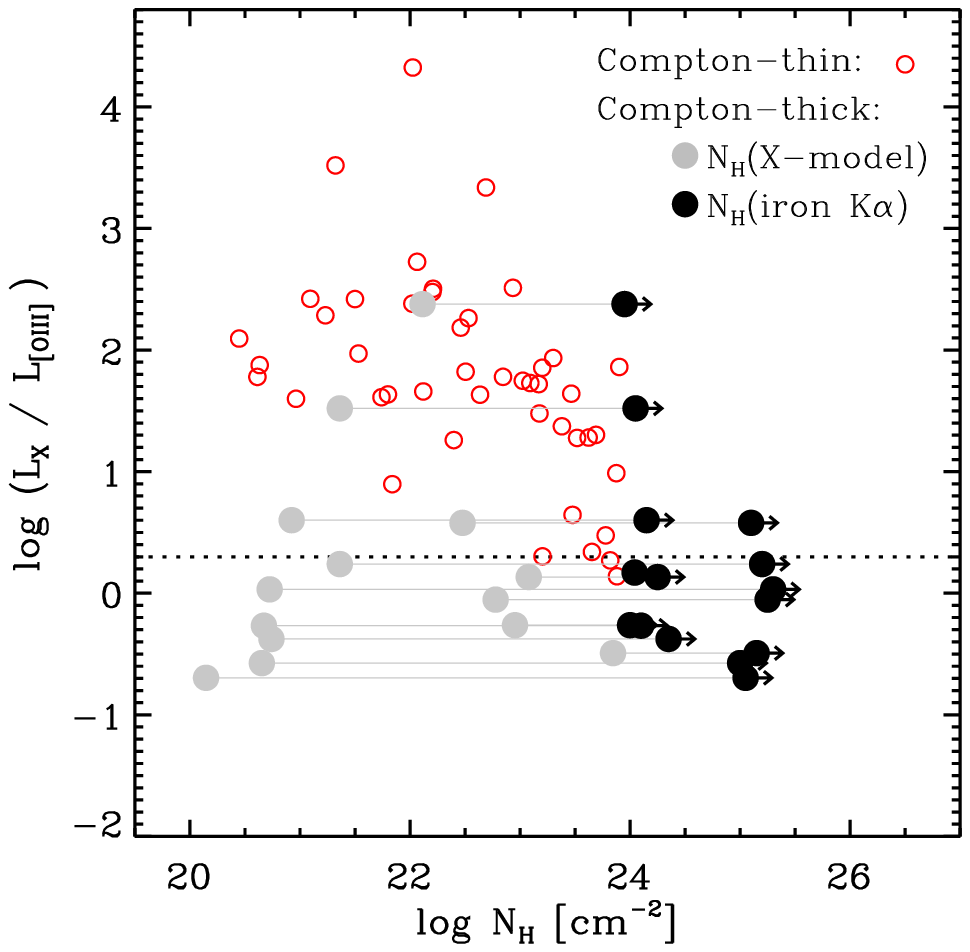}
\caption{
  The Compton-thickness parameter, $L_X(2-10$\,keV$)/L(\oiiilam)$, as a function of 
  X-ray absorption column density, $N_H$, for nearby galaxies in the sample from B99.  
  The column densities were derived by modeling the photoelectric cutoff in X-ray observations of Compton-thin (open circles) 
  and Compton-thick (filled gray circles) AGNs.  For the latter, a lower limit to $N_H$ is also provided from 
  the presence of Fe $K\alpha$ line with substantial equivalent width (filled black circles; B99) and/or 
  observations at energies beyond 10\,keV. The two estimates of $N_H$ for a given AGN are linked with thin solid lines. 
  The dotted line marks the adopted threshold for heavy X-ray absorption.
  Points with $N_H$ lower limits (of $N_H=10^{24}$ or $N_H=10^{25}$\,cm$^{-2}$) are shifted slightly horizontally 
  with respect to one another for clarity of the plotting symbols.
}\label{fig:Xabsor}
\end{figure}

As shown in Figure~\ref{fig:Xabsor}, the hard X-ray (2-10\,keV) to \oiiilam\ luminosity ratio 
drops when Compton-thickness 
is reached at $N_H\approx10^{24}$\,cm$^{-2}$ (filled black circles).  However, the 
most heavily-absorbed systems can be mistaken for low-column density cases with simple X-ray 
spectra models that do not account for additional signatures such as the equivalent 
width of the Fe $K\alpha$ emission line (which becomes high in cases of heavy absorption)  
or the absorption of even higher energy photons ($>$10\,keV).   
This feature is obvious in Figure~\ref{fig:Xabsor} 
where the two estimates of $N_H$ for Compton-thick AGNs (see B99 for details on $N_H$ calculations)
can differ by 5\,dex and typically by 3\,dex. This shows the failure of the photoelectric cutoff 
estimate of $N_H$ in cases of the heaviest absorption. 

The Compton-thickness parameter tends to be low for the most absorbed systems. 
We adopted the threshold log($L_X/L_{\oiii}$)$<$0.3\,dex to classify two AGNs that 
met the X-ray AGN criteria as X-absorbed AGNs rather than X-unabsorbed AGNs (Section~\ref{sec:agnsummary} and Table~1). 
However, it is possible that other AGNs in the X-unabsorbed category have in fact significant 
absorption as we do not use X-ray spectral signatures due to the low number of X-ray counts 
(and associated uncertainties) for AGNs in our sample.  In that sense, the absorbed fraction is a 
conservative lower limit and could be even higher if we moved more AGNs from the X-unabsorbed to 
the X-absorbed category.

%---------------------------------------------
%---------------------------------------------

In order to facilitate the comparison with the sample of interest in this work, we 
used the \oiiilam\ fluxes uncorrected for dust obscuration.  Applying the correction 
from the Balmer decrement method, as done by B99, would increase the \oiiilam\ fluxes 
by 1.6\,(1.0)\,dex on average (median) with some cases presenting more extreme obscuration (up to 2.7\,dex).  
In particular, two of the three Compton-thick AGNs with 
log($L_X/L_{\oiii}$)$>$0.3\,dex have extreme Balmer decrement values ($>$10), which 
implies that the observed \oiiilam\ line is underluminous and leads to a higher 
Compton-thickness parameter.  

%% END OF APPENDIX

\bibliographystyle{apj_sj}
\bibliography{AGNpaper_sj}

\end{document}